\documentclass[%
 aip,
 amsmath,amssymb,
 reprint,%
]{revtex4-2}\usepackage{amsmath}

\makeatletter
\newcommand*{\addFileDependency}[1]{
  \typeout{(#1)}
  \@addtofilelist{#1}
  \IfFileExists{#1}{}{\typeout{No file #1.}}
}
\makeatother

\usepackage{graphicx}
\usepackage{dcolumn}
\usepackage{bm}

\usepackage{graphicx}
\usepackage{dcolumn}
\usepackage{bm}

\usepackage[utf8]{inputenc}
\usepackage[T1]{fontenc}
\usepackage{mathptmx}
\usepackage{multirow}

\usepackage{xcolor}
\begin{document}

\preprint{AIP/123-QED}


\title{Band energy landscapes in twisted homobilayers of transition metal dichalcogenides}

\author{F. Ferreira}
\email{fabio.ferreira@postgrad.manchester.ac.uk}
\affiliation{Department of Physics \& Astronomy, University of Manchester, Manchester, M13 9PL, United Kingdom}
\affiliation{National Graphene Institute, University of Manchester, Booth St E, Manchester, M13 9PL, United Kingdom}
\author{S. J. Magorrian}
\affiliation{Department of Physics \& Astronomy, University of Manchester, Manchester, M13 9PL, United Kingdom}
\affiliation{National Graphene Institute, University of Manchester, Booth St E, Manchester, M13 9PL, United Kingdom}
\author{V. V. Enaldiev}
\affiliation{Department of Physics \& Astronomy, University of Manchester, Manchester, M13 9PL, United Kingdom}
\affiliation{National Graphene Institute, University of Manchester, Booth St E, Manchester, M13 9PL, United Kingdom}
\affiliation{Kotel'nikov Institute of Radio-engineering and Electronics of the Russian Academy of Sciences, 11-7 Mokhovaya St, Moscow, 125009 Russia}
\author{D. A. Ruiz-Tijerina}
\affiliation{Secretar\'ia Acad\'emica, Instituto de F\'isica, Universidad Nacional Aut\'onoma de M\'exico, Cd. de M\'exico C.P. 04510, M\'exico}

\author{V. I. Fal'ko}
\email{vladimir.falko@manchester.ac.uk}
\affiliation{Department of Physics \& Astronomy, University of Manchester, Manchester, M13 9PL, United Kingdom}
\affiliation{National Graphene Institute, University of Manchester, Booth St E, Manchester, M13 9PL, United Kingdom}
\affiliation{Henry Royce Institute for Advanced Materials, University of Manchester, Manchester, M13 9PL, United Kingdom}


\date{\today}

\begin{abstract}

Twistronic assembly of 2D materials employs the twist angle between adjacent layers as a tuning parameter for designing the electronic and optical properties of van der Waals heterostructures.
Here, we study how interlayer hybridization, weak ferroelectric charge transfer between layers, and piezoelectric response to deformations set the valence and conduction band edges across the moir{\'e} supercell in twistronic homobilayers of MoS$_2$, MoSe$_2$, WS$_2$ and WSe$_2$.
We show that, due to the lack of inversion symmetry in the monolayer crystals, bilayers with parallel (P) and anti-parallel (AP) unit cell orientations display contrasting behaviors.
For P-bilayers at small twist angles we find band edges in the middle of triangular domains of preferential stacking.  
In AP-bilayers at marginal twist angles ($\theta_{AP} < 1^\circ$) the band edges are located in small regions around the intersections of domain walls, giving highly localized quantum dot states.

\end{abstract}

\maketitle

\onecolumngrid
\twocolumngrid

The assembly of van der Waals heterostructures has potential for tailoring the properties of 2D materials\cite{Geim2013}. 
Recently, it has been shown that twisted bilayer graphene exhibits intrinsic unconventional superconductivity\cite{Cao2018} and a Mott insulating phase\cite{2Cao2018}. 
These effects have been related to the localization of electrons in particular stacking areas of the moir\'e superlattice (mSL), potentially enhanced by lattice reconstruction, promoting energetically favorable Bernal stacking\cite{Woods2014,ZHANG2018225,PhysRevB.95.115429}.

Lattice reconstruction also takes place in bilayers of twisted transition metal dichalcogenides\cite{carr2018relaxation,PRLNaik,rosenberger2020,PhysRevLett.124.206101,Weston2020,McGilly2020,Edelberg2020,Wang2020,abdd92,vitale2021flat}  (TMDs), where it gives rise to the formation of preferential stacking domains separated by a network of domain walls (DWs). 
The shape and size of these domains depend on the mutual orientation of the unit cells of the individual crystals, which can be parallel (P) or anti-parallel (AP), whereas the superlattice period depends on the misalignment angle between the crystallographic axes of the layers. 
These two orientations fundamentally differ in that AP-bilayers possess inversion symmetry for all local stacking configurations, whereas for P-bilayers both inversion and mirror reflection symmetries are generally broken.
Below we discuss how this symmetry breaking in P-bilayers leads to:
(a) interlayer charge transfer due to hybridization of the conduction bands in one layer with the valence bands of the other; 
(b) layer-asymmetric piezoelectric charges caused by lattice reconstruction, concentrated around the network of domain walls and corners.
Some of these effects were recently discussed in relation to the structural characterization of TMD homobilayers and heterobilayers\cite{carr2018relaxation,PRLNaik,rosenberger2020,PhysRevLett.124.206101,Weston2020,McGilly2020,Edelberg2020,Wang2020,abdd92,vitale2021flat}, and in twisted hexagonal boron nitride (hBN)\cite{Woods2021,2020arXiv201006600Y,walet2020flat}.
In both P- and AP-bilayers, the band energies depend on the interplay between the above effects and the interlayer hybridization of band edge states, which is most prominent for the states around the $\Gamma$- and Q-valleys and marginally relevant for the K-valley band edges.

To model twisted TMD homobilayers, we use a multiscale  approach\cite{PhysRevLett.124.206101} for describing lattice reconstruction and computing piezoelectric charge and potential distributions, complemented by DFT-parametrized Hamiltonians for interlayer hybridization of band edge states developed to describe an arbitrary local stacking.
The first step is to use an earlier  parametrized\cite{PhysRevLett.124.206101} interlayer adhesion potential to compute  atomic reconstruction of twisted TMDs for various twist-angles using elasticity theory. We take into account both lateral and vertical deformations of the layers, see S1 in Supplementary Material (SM).
This gives us the pattern of local interlayer distance $d$ and  lateral offset $\bm{r}_0$, which determine the stacking of the layers, as well as the piezoelectric potential distribution across the mSL.  
The second step is to compute the band structures of bilayers with various offsets (using DFT implemented in Quantum ESPRESSO package\cite{QEcite}) in order to parametrize 
$\bm{r}_0$-dependent Hamiltonians describing interlayer hybridization (see S2 and S3 in SM). Diagonalizing the Hamiltonians,  which take into account piezopotentials, we build maps for the local band edge energies across the mSL. 

We analyze the band edge properties in the valence band (VB), near both the $\Gamma$- and K-valleys, and the conduction band (CB), near both the K- and Q-valleys.
We consider both options ($\Gamma$ and K) for the location of the VB edge in the Brillouin zone, as their relative energies may depend on the encapsulating environment. In the presence of hBN, the energies of chalcogen orbitals, which are strongly represented at the $\Gamma$-valley, may be shifted with respect to the metal $d_{x^2-y^2}$ and $d_{xy}$ orbitals that determine the K-valley energy\cite{Korm_nyos_2015}, with the magnitude and direction of the shift determined by the WSe$_2$/hBN band alignment, and the properties of their hybridization.

Local stacking configurations, distinguished by relative lateral offset of the two lattices in the bilayer (see Fig. \ref{fig:stackings_supl} in SM), vary across the moir\'e supercell, with lattice reconstruction promoting energetically favorable configurations. For P-stacking, these are MX$'$/XM$'$, in which a metal atom M(M$'$)  is below(above) a chalcogen atom X(X$'$) in the opposite layer (X and X$'$ refer to chalcogen atoms in bottom and top layer; M and M$'$ refer to transition metal atoms in bottom and top layer). For AP-stacking, the favorable configuration is 2H, as found in bulk TMD crystals, and features two interlayer pairs of vertically aligned metal/chalcogen atoms. Other important high-symmetry stackings include XX$'$ (both P and AP), for which chalcogen atoms are vertically aligned, and for AP stacking MM$'$, which has aligned metal atoms (Fig. \ref{fig:stackings_supl} in SM).
P-bilayer superlattices are comprised of two triangular domains with MX$'$ and XM$'$ stacking separated by domain walls with XX$'$ stacking near hexagonal domain wall lattice sites, whereas in AP-bilayers there are large 2H domains separated by a honeycomb network of domain walls with XX$'$ and MM$'$ stacking at their intersections\cite{PhysRevLett.124.206101}. 
Below, we aim to determine which stacking areas in the moir\'e supercell host the band edges, which is essential for determining and predicting the formation of quantum dots in marginally twisted structures. We discuss separately features of the K-, Q- and $\Gamma$-valley band edges for P- and AP-bilayers.

\onecolumngrid

\begin{figure}[h]
\includegraphics[scale=0.24]{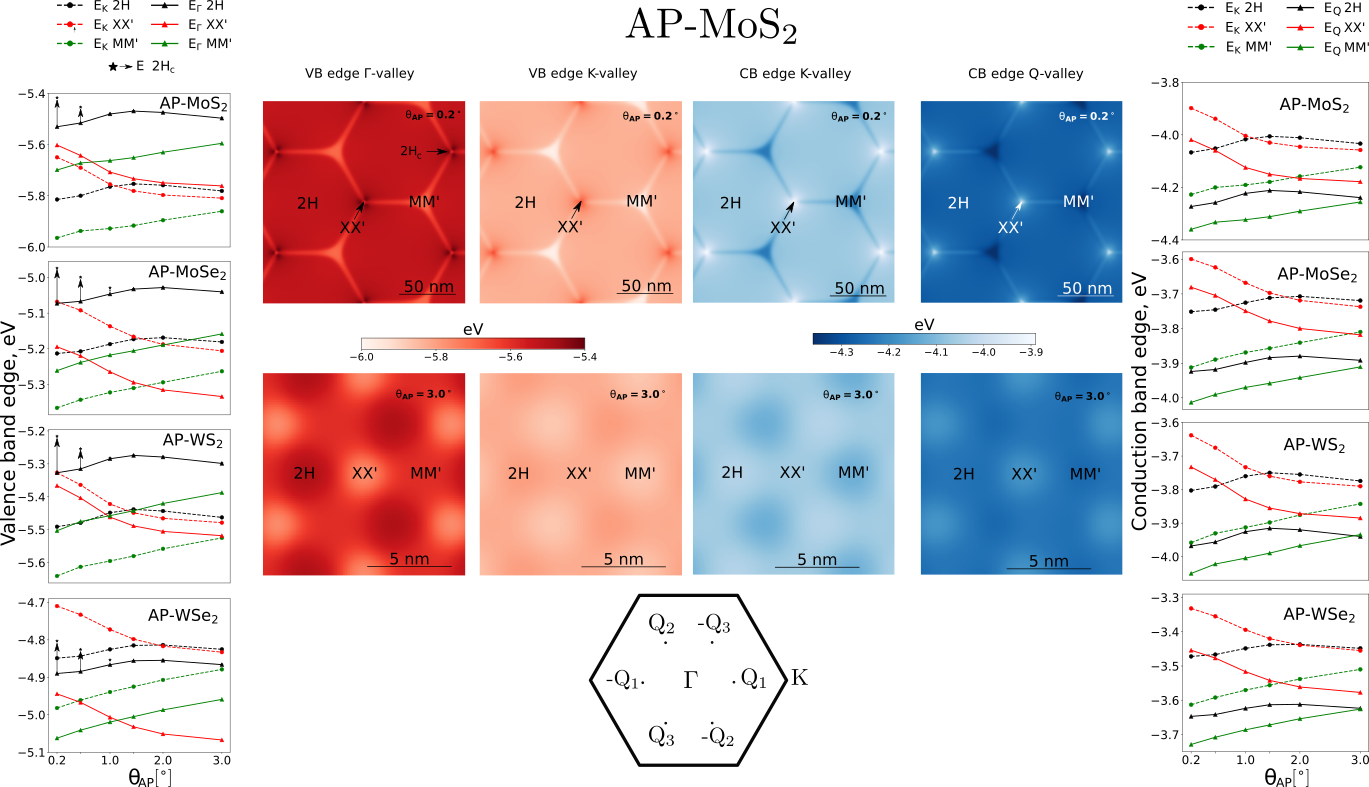}
\caption{
Left panels: Variation of the VB energy with twist angle $\theta_{AP}$ for different stacking configurations in AP-bilayers at the $\Gamma$- and K-valleys. 
The VB edge at the $\Gamma$-valley is located at the corners of 2H domains (labeled as 2H$_{\rm c}$) for marginal twist angles.
These corners can be seen in the $\Gamma$-valley VB edge map for $\theta_{AP} = 0.2^\circ$.
The arrows in the left panels mark the difference between the $\Gamma$-valley VB energy at the corners and in the center of 2H domains.
Right panels: Variation of the CB energy with $\theta_{AP}$ for different stacking configurations in AP-bilayers at the K- and Q-valleys. 
The middle panels show maps for the VB edge at the $\Gamma$- and K-valleys, and for the CB edge at the K- and $\pm$Q$_{1}$-valleys in AP-MoS$_2$ for $\theta_{AP} = 0.2^\circ$ and $\theta_{AP} = 3^\circ$.
The zigzag orientations of DWs make the Q-valley CB edge maps $C_3$-symmetric, leading to the same CB edge landscapes for $\pm$Q$_{1,2,3}$.
The vacuum level is set to 0~eV.
See  Figs. \ref{fig:APMoS2_band_maps_supl}-\ref{fig:APWS2_band_maps_supl} with maps for the other TMDs. }
\label{fig:APMoS2_valence_band_maps}
\end{figure}   

\twocolumngrid

\begin{figure*}
\includegraphics[scale=0.24]{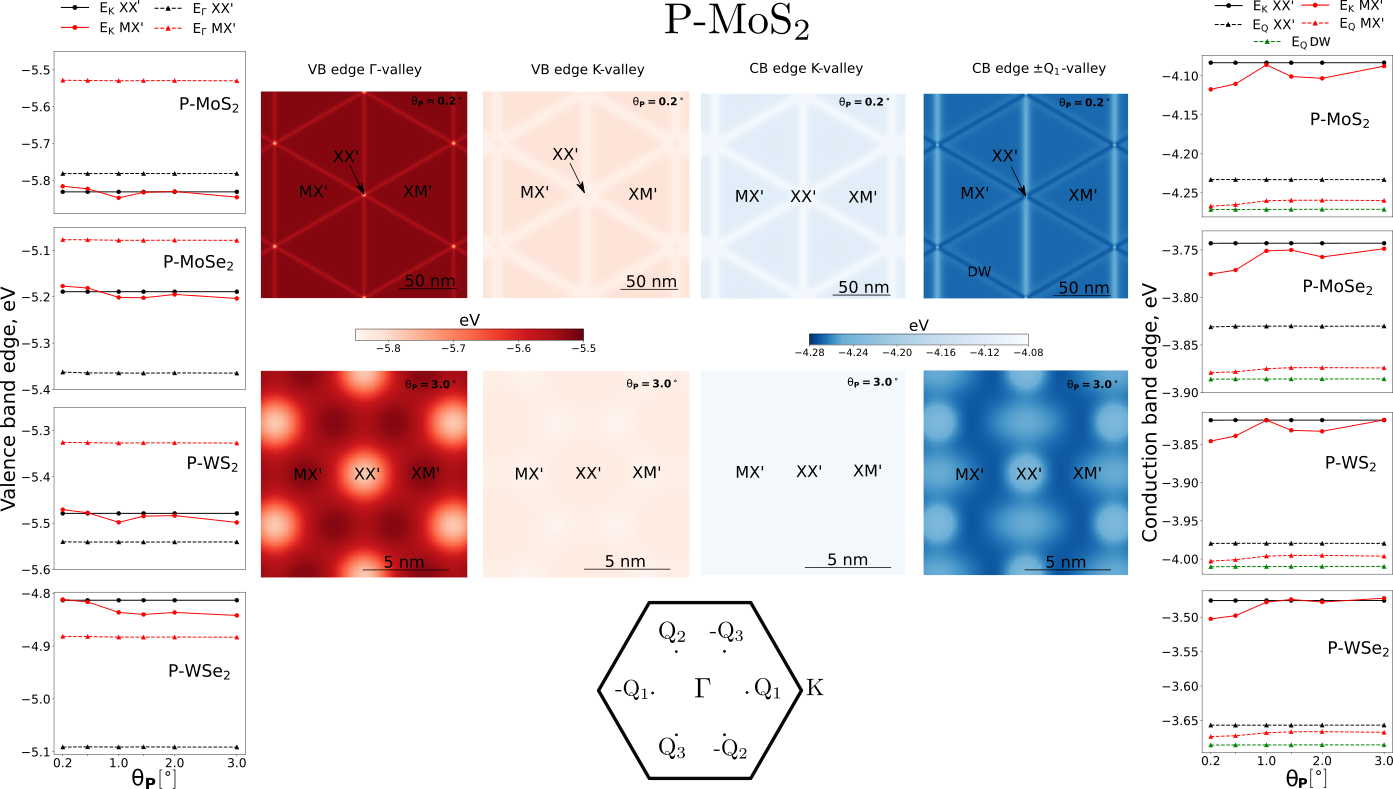}
\caption{
Left panels: Variation of the VB energy with twist angle $\theta_P$ for XX$'$- and MX$'$-stacking configurations in P-bilayers at the $\Gamma$- and K-valleys. 
Right panels: Variation of the CB energy with $\theta_P$ for XX$'$- and MX$'$-stacking configurations in P-bilayers at the K- and Q-valleys. 
The middle panels show maps for the VB edge at the $\Gamma$- and K-valleys, and for the CB edge at the K- and $\pm$Q$_{1}$-valleys in P-MoS$_2$ for $\theta_{P} = 0.2^\circ$ and $\theta_{P} = 3^\circ$. 
Note that for all angles the Q-valley maps are anisotropic, with the anisotropy axis rotated by $\pm120^{\circ}$ when going from Q$_1$ to Q$_{2}$ and Q$_3$. For  $\theta_{P} = 3^\circ$, the low contrast of the CB and VB K-valley maps  reflect a negligible variation of the band edge energy ($< 20 $ meV).  The vacuum level is set to 0 eV. See Figs.\ \ref{fig:PMoS2_band_maps_supl}-\ref{fig:PWS2_band_maps_supl} for maps for the other TMDs.}
\label{fig:PMoS2_valence_band_maps}
\end{figure*}

\textbf{\textit{For AP-bilayers the position of the VB at $\Gamma$}} is determined by the interplay between the piezopotential and interlayer hybridization of resonant band edges.
Due to symmetry, the piezopotential induced by lateral lattice reconstruction is the same on both layers\cite{PhysRevLett.124.206101}; interlayer hybridization is also sensitive to lattice deformations, but in the vertical direction.
This is because the wavefunctions at $\Gamma$
carry a substantial weight of $p_z$ orbitals of chalcogens, which strongly overlap between the layers with a pronounced dependence on the interlayer distance\cite{Korm_nyos_2015}.

2H domains, which provide energetically more favorable stacking, also feature the closest interlayer distance, promoting the higher position of the top VB due to interlayer hybridization of the VB edges.
At the same time, piezocharges are largest and negative (hence, the electron piezopotential is highest) at XX$'$ regions\cite{PhysRevLett.124.206101}, attracting holes from the centers of 2H domains toward XX$'$ corners.
The trigonal symmetry of the honeycomb domain structure suggests that the resulting VB edges appear as three maxima labeled in Fig.\ \ref{fig:APMoS2_valence_band_maps} as 2H$_{\rm c}$.
This behavior is corroborated by the plots shown on the left hand side of Fig.\ \ref{fig:APMoS2_valence_band_maps}, where band edge energies in areas with local 2H, MM$'$ and XX$'$ stacking configurations are compared with each other, and with the energy of those local maxima at 2H$_{\rm c}$.
The data shown in Fig.\  \ref{fig:APMoS2_valence_band_maps} for all TMDs indicate that they all display the same VB behavior at $\Gamma$.

For large twist angles, the piezopotential magnitude decreases, whereas the contribution of interlayer hybridization remains the same.
This promotes the maximum value for the VB energy at the center of 2H domains, while the minimum value switches from MM$'$- to XX$'$-stacking regions. 
This is shown in the left-hand-side panels of Fig.\ \ref{fig:APMoS2_valence_band_maps}, where the crossover between the two regimes takes place at $\theta_{AP}\approx 0.6^\circ$.

\textbf{\textit{For AP-bilayers the VB and CB edge energies at K}} are dominated by the piezopotential contribution.
This is because interlayer hybridization between K-valley states in  AP-bilayers is weak:  the band edge states are dominated by the orbitals of metals, buried inside the layer\cite{Korm_nyos_2015}, and their   spin-valley  locking due to spin-orbit splitting makes monolayer bands with the same spin off-resonant.   

A domain wall structure is fully developed for twist angles $\theta_{AP} \lesssim 0.6^\circ$, giving rise to large piezopotentials around domain wall intersections. For such twist angles, the piezopotential forms quantum-dot-like wells of depth $\sim$150~meV for electrons and holes in the MM$'$ and XX$'$ intersections of the domain wall structure, respectively, shown in Fig.\ \ref{fig:APMoS2_valence_band_maps}. 
In a periodic moir\'e supercell, quantized states localized in each of these quantum dots should give rise to narrow bands located at the CB and VB edges of marginally twisted bilayers.
Upon increasing the twist angle, the magnitude of the piezopotential decreases and the location of its minimum in the supercell shifts to 2H areas, followed by the position of the VB band edge. The CB edge remains in the MM$'$-stacked regions due to a residual attractive piezopotential for electrons even at larger angles. 

\textit{\textbf{For AP-bilayers the position of the CB edge at the six Q-valleys}} is determined by the piezopotential and by interlayer coupling variations across the mSL. For $\theta_{AP}\leq 1^{\circ}$, the piezopotential overcomes hybridization, forming quantum dots for electrons at MM$'$ corners of 2H domains.  For $\theta_{AP}>3^{\circ}$, the Q-valley CB edge shifts to the center of  2H regions. 

\textbf{\textit{For P-bilayers the VB edge at $\Gamma$}} is dominated by interlayer hybridization of chalcogen orbitals, which is an order of magnitude stronger than both the piezoelectric and ferroelectric potentials (Table \ref{tab:fittings_GAMMA} in  SM). 
A weak ferroelectric potential arises due to the interlayer charge transfer caused by off-resonant hybridization of conduction bands in one layer with valence bands in the opposite layer. Such a spontaneous out-of-plane charge polarisation is allowed due to the lack of inversion symmetry in P-bilayers \cite{Li2017,Tong_2020}.
For the full range of twist angles, the VB at $\Gamma$  lies in the MX$'$/XM$'$ domains, where hybridization is largest because the interlayer distance is smallest (see Fig. \ref{fig:PMoS2_valence_band_maps}).

\textbf{\textit{For P-bilayers the VB and CB edges at K}} are determined by a competition between interlayer charge transfer, the piezopotential, and resonant interlayer hybridization at the band edges. All three are comparable for the K-valley states, but vary differently across the mSL. Interlayer hybridization vanishes for MX$'$ and XM$'$ stackings due to the symmetry of the K-valley Bloch states in the honeycomb lattice\cite{PhysRevB.99.125424}, whereas the potential step due to ferroelectric charge transfer is largest (Tables \ref{tab:fittings_GAMMA} and \ref{tab:fittings_K} in  SM).
The piezopotential, which has opposite signs in the two layers, contributes to the splitting of the resonantly coupled bands rather than to the modulation of their average position, as in AP-bilayers\footnote{Unlike the case of AP-bilayers, in P-bilayers the top and bottom layers have piezocharges of opposite sign. 
This is due to the signs of the piezocoefficients in the expression for piezocharge density. See the supplementary material in Ref.[\onlinecite{PhysRevLett.124.206101}]}. 
Splitting in XX$'$ areas is only determined by interlayer hybridization, as the ferroelectric potential and piezopotential are absent there.  The  piezopotential distribution  strongly depends on twist angle. Also, piezo- and ferroelectric potentials have opposite signs, and fully or partially cancel each other out in areas where both are present. 
At marginal twist angles $\theta_P\lesssim 1^{\circ}$, this leads to flat  CB and VB edges across  MX$'$ and XM$'$ domains, see Fig. \ref{fig:PMoS2_valence_band_maps}, forming shallow ($\approx 30$\,meV) triangular traps for electrons and holes, respectively. 

As the twist angle is increased, the piezopotential expands into the MX$'$/XM$'$ domains, with similar magnitude and opposite sign to the potential caused by the charge transfer effect, reducing the splitting between the bands (SM Fig.\ \ref{fig:P_piezo_maps_supl}). For the VB, this is sufficient to move the band edge to the XX$'$ areas, where the splitting due to weak interlayer hybridization persists, and for the CB the MX$'$/XM$'$ and XX$'$ regions become very close in energy, see right hand panels of Fig. \ref{fig:PMoS2_valence_band_maps}.

\textit{\textbf{For P-bilayers, the Q-valley CB edge}} is mainly affected by the  variation of resonant interlayer hybridization of monolayer Q-valley states across the mSL, with only weak effects from the layer-asymmetric piezo- and ferroelectric  potentials. For $\theta_P\leq 2^{\circ}$, the CB edge  appears at one-dimensional channels zig-zagging across the mSL. Such channels are $\sim 25$ meV deep for P-WSe$_2$, but shallower for the other TMDS. The asymmetry seen in the 
band edge maps is due to the underlying asymmetry of the Q-valley wave functions in each layer, interplaying with the lateral interlayer offset, which is differently oriented at different segments of the DW web. 
 In this regime, XX$'$ areas play the role of potential barriers, acting as scatterers between the channels for CB electrons. For $\theta_P\geq 2^{\circ}$, the channels broaden to form  anisotropic landscapes. Despite their small amplitude, ``potential'' variations in such landscapes would determine anisotropic moir\'e minibands for Q-valley electrons.

\textit{\textbf{Finally}}, we have established how the interplay between 
interlayer hybridization, piezoelectric potential and ferroelectric charge transfer affects the band edges for electrons in the K- and Q-valleys in the CB, and K- and $\Gamma$-valleys in the VB of TMD bilayers. 
We have analyzed all these possibilities because the influence of the encapsulating  material or a substrate may affect their mutual alignment. 
For example, the DFT modelling of isolated bilayers points towards Q-valley CB edge for all four TMDs studied here, whereas recent experimental studies \cite{Enslin_PhysRevResearch} of 2H MoS$_2$ bi- and tri-layers indicate the persistence of K-valley band edges. 
Then, the crossovers between different  misalignment angle regimes presented in this paper and highlighted in Figs.\ \ref{fig:APMoS2_valence_band_maps} and \ref{fig:PMoS2_valence_band_maps} represent the variety of scenarios for the formation of quantum dot arrays that determines Hubbard physics for narrow minibands of electrons and holes in twistronic bilayers with different encapsulation environments.

\section*{Supplementary Material}

See the Supplementary Material for more details about the multiscale approach for  describing  lattice  reconstruction and computing piezoelectric charge and potential distributions.
More details about DFT bandstructure calculations and effective Hamiltonians used to describe band edges at $\Gamma$-, K and Q-valleys can also be seen in the Supplementary Material.

\section*{Acknowledgements}

We thank C. Yelgel, N. Walet, Q. Tong, M. Chen, F. Xiao, H. Yu and W. Yao for discussions.
This work has been supported by EPSRC grants EP/S019367/1, EP/V007033/1, EP/S030719/1, EP/N010345/1; ERC Synergy Grant Hetero2D; Lloyd’s Register Foundation Nanotechnology grant; European Graphene Flagship Project, and EU Quantum Technology Flagship project 2D-SIPC. Computational resources were provided by the Computational Shared Facility of the University of Manchester and the ARCHER2 UK National Supercomputing Service (https://www.archer2.ac.uk) through EPSRC Access to HPC project e672.

\section*{DATA AVAILABILITY}
The data that support the findings of this study are available from the corresponding author upon reasonable request.

\section*{REFERENCES}
\bibliography{bibl}

\clearpage

\onecolumngrid

\begin{center}
{\large Supplementary material for: Band energy landscapes in twisted homobilayers of transition metal dichalcogenides}
\end{center}

\renewcommand{\thesection}{S\arabic{section}}  
\renewcommand{\thetable}{S\arabic{table}}  
\renewcommand{\thefigure}{S\arabic{figure}} 
\def\theequation{S\arabic{equation}}
\setcounter{figure}{0}

\section{Lattice reconstruction and piezopotential}

\begin{figure}[h]
\includegraphics[scale=0.2]{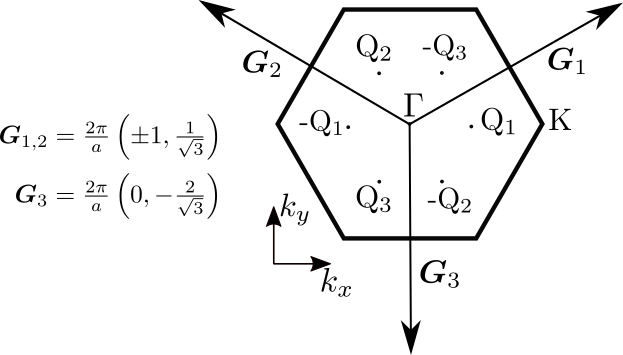}
\caption{ Reciprocal vectors  $\bm{G}_{1,2,3}$ and location of  $\Gamma$-, K- and Q-valleys in the Brillouin zone.   }
\label{fig:BZ}
\end{figure}

\begin{figure}[h]
\includegraphics[scale=0.07]{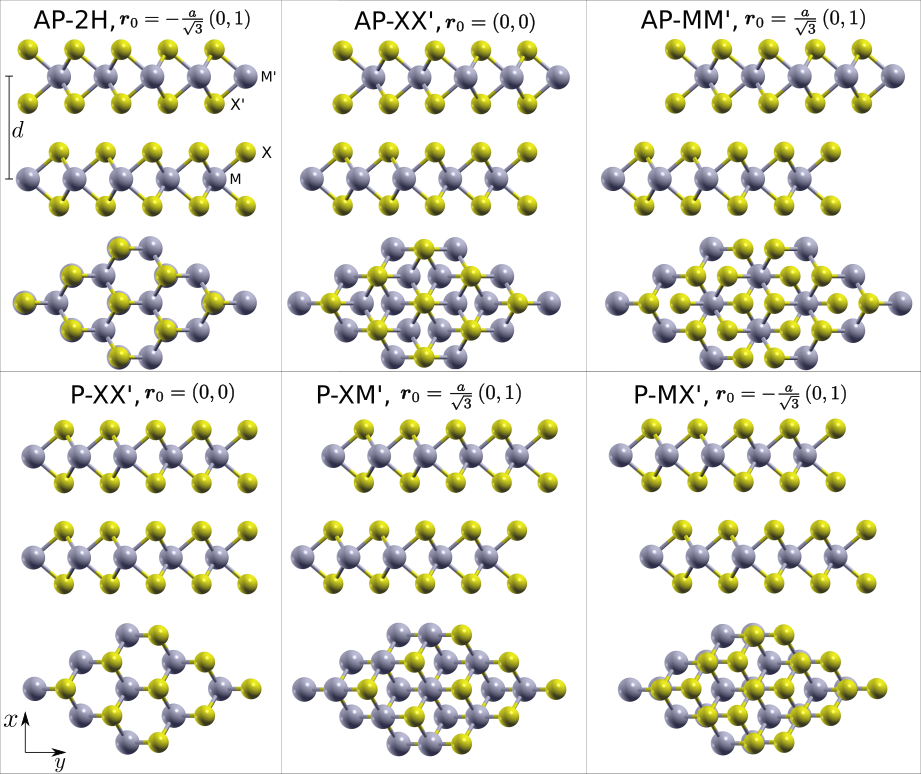}
\caption{Stacking configurations for AP- and P-bilayers where M(X) represents the metal(chacolgen) atoms in the bottom layer and M'(X') represents the metal(chacolgen) in the top layer. The quantity $d$ is the interlayer distance, $\boldsymbol{r}_0$ is the in-plane relative shift of the layers and $a$ is the lattice constant listed in Table \ref{tab_Fit}. 
AP-2H: M and X' atoms are vertically aligned with a local shift of $\boldsymbol{r}_0=-a/ \sqrt{3} (0,1)$. 
AP-XX$'$: X and X$'$ atoms are vertically aligned with a local shift of $\boldsymbol{r}_0=(0,0)$.
AP-MM$'$: M and M$'$ atoms are vertically aligned with a local shift of $\boldsymbol{r}_0=a/ \sqrt{3} (0,1)$.
P-XX$'$: X and X$'$ atoms are vertically aligned with a local shift of $\boldsymbol{r}_0=(0,0)$.
P-MX$'$: M and X$'$ atoms are vertically aligned with a local shift of  $\boldsymbol{r}_0 -=a/ \sqrt{3} (0,1)$.
P-XM$'$: X and M$'$ atoms are vertically aligned with a local shift of  $\boldsymbol{r}_0=a/ \sqrt{3} (0,1)$.   }
\label{fig:stackings_supl}
\end{figure}

{\bf Modeling of lattice relaxation in twisted TMD bilayers}. We describe lattice reconstruction in twisted bilayers in frames of mesoscale elasticity theory supplemented by DFT-parametrized adhesion energy derived in Refs. \cite{abdd92,PhysRevLett.124.206101,Weston2020}. In this approach we introduce in-plane deformation fields $\bm{u}^t$ and $\bm{u}^b$, which ensure local minimum for sum of elastic and adhesion energies in the moir\'e supercell:
\begin{equation}\label{Eq:elasticity}
    \mathcal{E}=\int_{\rm supercell}d^2\bm{r}\left\{\sum_{l={t,b}}\left[\frac{\lambda_l}{2}\left(u_{ii}^{l}\right)^{2} + \mu_l u_{ij}^{l}u_{ji}^{l}\right] + W_{P/AP}(\bm{r}_0(\bm{r}))\right\}.
\end{equation}
Here, $\lambda_{t/b}$, $\mu_{t/b}$ are elastic moduli of TMD monolayers (listed in Table \ref{tab_Fit}), $u_{ij}^{t/b}=\frac{1}{2}(\partial_ju_{i}^{t/b}+\partial_iu_{j}^{t/b})$ is top/bottom layer in-plane strain tensor $i,j=x,y$, and $W_{P/AP}(\bm{r}_0(\bm{r}))$ is the adhesion energy taken at local offset between the twisted layers\cite{PhysRevLett.124.206101,abdd92} 
\begin{equation}\label{eq:local_shiftr0}
\boldsymbol{r}_{0}(\boldsymbol{r})=\theta \hat{z} \times \boldsymbol{r}+\boldsymbol{u}^{t}-\boldsymbol{u}^{b}.
\end{equation}
Reference of the local offsets $\bm{r}_0(\bm{0})=(0,0)$ in Eq. (\ref{eq:local_shiftr0}) corresponds to XX$'$ configuration for P/AP-orientation of bilayers shown in Fig. \ref{fig:stackings_supl}. Explicit form for the adhesion energy reads
\begin{align}\label{Eq:adhesion_2}
    W_{P/AP}(\bm{r}_0) = -\kappa Z^2(\bm{r}_0) + 
    \sum_{n=1}^{3}\left[w_1\cos\left(\bm{G}_n\bm{r}_0\right) + w_2\sin\left(\bm{G}_n\bm{r}_0+\gamma_{P/AP}\right)\right], \\
    Z(\bm{r}_0)=\frac{1}{2\kappa}\sum_{n=1}^3\left[ w_1Q\cos\left(\bm{G}_n\bm{r}_0\right) + w_2G\sin\left(\bm{G}_n\bm{r}_0+\gamma_{P/AP}\right)\right], \nonumber
\end{align}
where $\gamma_{AP}=0$, $\gamma_P=\pi/2$, $\bm{G}_{1,2,3}$ are the shortest reciprocal vectors of TMD monolayer with magnitude $G$ related by $C_3$-rotation (see Fig. \ref{fig:BZ}), the term $-\kappa Z^2(\bm{r}_0)$ takes into account relaxation of interlayer distances $d_0+Z(\bm{r}_0)$ with stacking configurations as it was derived in Ref. \cite{PhysRevLett.124.206101} ($d_0$ is optimal distance for stacking-averaged adhesion energy). Values of parameters in adhesion energy (\ref{Eq:adhesion_2}) for studied homobilayers are listed in Table \ref{tab_Fit}.

{\setlength{\tabcolsep}{3.2pt}
\renewcommand{\arraystretch}{1.2}
\begin{table}[!h]
	\caption{Elastic moduli\cite{iguiniz2019,androulidakis2018} and piezocoefficients\cite{zhu2015,rostami2018piezoelectricity} for TMD monolayers, and parameters for the adhesion energy density  (\ref{Eq:adhesion_2}) of corresponding homobilayers. $a$ is the lattice constant. \label{tab_Fit}}
	\begin{tabular}{c|ccccccccc}
		\hline
		\hline
		& $w_1$, & $w_2$, & $\kappa$,  & $Q$, & $e_{11}$, & $\mu$ ,& $\lambda$, & $d_0$, & $a$,\\ 
		&\mbox{eV$\cdot$nm$^{-2}$} &\mbox{eV$\cdot$nm$^{-2}$} &  \mbox{eV$\cdot$nm$^{-4}$} &  nm$^{-1}$ & $10^{-10}$  C$\cdot$m$^{-1}$ & N/m & N/m & nm  & nm\\
		\hline 
		MoS$_2$ & 0.1727 &  0.0186 & 214 & 30.534 & 2.9 & 70.9 & 83.2 &0.65&  0.316  \\ 
		WS$_2$ &  0.1625 &  0.0224 & 213 & 30.877 & 2.74 & 72.5 & 52.5 &0.65&  0.315 \\
		MoSe$_2$ &0.1725  & 0.0250 & 189 & 29.614 & 2.14 & 49.6 &  42.3 &0.68& 0.329 \\
		WSe$_2$ & 0.1340 &  0.0261 & 190 & 29.889 & 2.03 & 48.4 & 29.7 & 0.69& 0.328 \\
		\hline
		\hline
	\end{tabular}
\end{table}
}

Minimization of the functional (\ref{Eq:elasticity}) results in a system of four Euler-Lagrange equations that are discretized on a rectangular grid  using the finite-difference method. 
The package GEKKO Optimization Suite\cite{gekko_package} is used to solve this system of non-linear equations, in which the final solution produces a detailed structure of the strain fields. 

{\bf Piezopotential}. Lacking inversion center, TMD crystals become piezoactive in the limit of a single monolayer \cite{zhu2015}. Therefore, deformations in each layer induced by lattice relaxation result in the emergence of piezoelectric charge densities in the top/bottom layers:
\begin{equation}
\rho^{t/b} = -e_{11}^{t / b}\left[2 \partial_{x} u_{x y}^{t / b}+\partial_{y}\left(u_{x x}^{t / b}-u_{y y}^{t / b}\right)\right],
\end{equation}
where $e_{11}^{t/b}$ are piezocoefficients listed in Table \ref{tab_Fit}. The latter are the same ($e_{11}^t = e_{11}^b>0$) for P- and opposite ($e_{11}^b=-e_{11}^t>0$) for AP-bilayers. We note that opposite signs of the deformation fields in the two layers $\bm{u}^t=-\bm{u}^b$ lead to the same sign of the piezocharges in the constituent layers for AP-bilayers and opposite signs of piezocharges for P-bilayers. To calculate the potential $\phi^{t/b}$ created by the piezocharges in the top/bottom layers we expand them in Fourier series over the moir\'e superlattice reciprocal vectors solving Poisson equation with appropriate boundary conditions on interfaces taking into account in-plane polarisation of orbitals and hBN encapsulation as it is described in Ref. \cite{abdd92}.

Lattice reconstruction results from the competition between elastic and adhesion forces. 
This is expressed in the following functional\cite{PhysRevLett.124.206101}.
\begin{equation}
\mathcal{E}=\int_{\mathrm{supercell}} d^{2} \boldsymbol{r}\left(U+\widetilde{W}_{A P/P}\left(\boldsymbol{r}_{0}(\boldsymbol{r}), d(\boldsymbol{r})\right)\right),
\label{eq:functional}
\end{equation}
where $\boldsymbol{r}_{0}$ is the local shift between chalcogen atoms in each layer, $d$ is the interlayer distance,  $U$ is the elastic energy and $\widetilde{W}_{AP/P}$ is the local adhesion energy that takes into account in-plane and out-plane reconstruction.
The expression for the elastic energy is parametrised as follows 
\begin{equation}
U=\sum_{l=t, b}\left[\left(\lambda_{l} / 2\right)\left(u_{i i}^{(l)}\right)^{2}+\mu_{l}\left(u_{i j}^{(l)}\right)^{2}\right]
\end{equation}
where  $\lambda_{t/b}$ is first Lamé coefficients for the top ($t$) and  bottom ($b$) layer, $\mu_{t/b}$ is the shear modulus and $u_{i j}^{(t / b)}=\frac{1}{2}\left(\partial_{j} u_{i}^{(t / b)}+\partial_{i} u_{j}^{(t / b)}\right)$ is the strain tensor.
The formula for $\widetilde{W}_{A P/P}$ is expressed as follows
\begin{equation}
\begin{aligned}
\widetilde{W}_{A P/P}=& \left\{\varepsilon Z_{\mathrm{P} / \mathrm{AP}}^{2}\right.\\
&+\sum_{n=1}^{3}\left[A_{1} e^{-\sqrt{G^{2}+\rho^{-2}} d_{0}} \cos \left(\boldsymbol{g}_{n} \boldsymbol{r}_{0}+\boldsymbol{G}_{n}\left[\boldsymbol{u}^{(t)}-\boldsymbol{u}^{(b)}\right]\right)\right.\\
&\left.\left.+A_{2} e^{-G d_{0}} \sin \left(\boldsymbol{g}_{n} \boldsymbol{r}_{0}+\boldsymbol{G}_{n}\left[\boldsymbol{u}^{(t)}-\boldsymbol{u}^{(b)}\right]+\varphi_{\mathrm{P} / \mathrm{AP}}\right)\right]\right\}, \\
Z_{\mathrm{P} / \mathrm{AP}}=&\left.\frac{1}{2 \varepsilon} \frac{\partial}{\partial d}\left[f(d)-W_{\mathrm{P} / \mathrm{AP}}(\boldsymbol{r}_{0}, d)\right]\right|_{d=d_{0}},\\
W_{\mathrm{P} / \mathrm{AP}}\left(\boldsymbol{r}_{0}, d\right)=& \sum_{n=1}^{3}\left[-\frac{C_{n}}{d^{4 n}}+A_{1} e^{-\sqrt{G^{2}+\rho^{-2}} d} \cos \left(\boldsymbol{G}_{n} \boldsymbol{r}_{0}\right) +A_{2} e^{-G d} \sin \left(\boldsymbol{G}_{n} \boldsymbol{r}_{0}+\varphi_{\mathrm{P} / \mathrm{AP}}\right)\right],
\end{aligned}
\end{equation}
where $\boldsymbol{g}_{n}$ are the reciprocal vectors of the moir\'e supercell, $\boldsymbol{G}_{n}$ are the reciprocal vectors of TMD where $G$ is their magnitude, $\mathbf{u}^{t/b} =\left(u_{x}^{t/b},u_{y}^{t/b}\right)$ are the displacement field vectors, $d_0$ is the minimum value for interlayer distance, $f(d)\approx f(d_0) +\varepsilon \left(d-d_0\right)^2$ is stacking-averaged term and $\varphi_{P} = \pi/2$ and $\varphi_{AP} = 0$, 
The coefficients $A_{1,2},\rho,\epsilon$, $C_{1,2,3}$ are obtained by fitting the adhesion energy to ab initio DFT calculations as explained in more detail in Ref. \onlinecite{PhysRevLett.124.206101}.
The reconstructed local shift between chalcogen atoms is given by
\begin{equation}\label{eq:local_r0}
\boldsymbol{r}_{0}(\boldsymbol{r})=\theta \hat{z} \times \boldsymbol{r}+\delta \boldsymbol{r}+\boldsymbol{u}^{(t)}-\boldsymbol{u}^{(b)},
\end{equation}
where $\delta$ is the lattice mismatch between two layers and $\theta \hat{z}$ is a rotation that results from the misalignment angle.
The displacement vectors can be obtained by minimizing Eq. (\ref{eq:functional}) with respect to $\mathbf{u}^{t/b}$. 
This minimization results in a system of Euler-Lagrange equations and it is solved using finite difference method with the help of the package GEKKO Optimization Suite \cite{Beal_2018}. 
Lattice reconstruction will also influence the rearrangement of the charges in this TMDs. 
Thus, piezoelectricity plays an important role on these materials given that they lack inversion symmetry.
The expression for bare piezocharges is given by\cite{PhysRevLett.124.206101}
\begin{equation}
\rho^{t/b}(\boldsymbol{r},z) =  -e_{11}^{t / b}\left[2 \partial_{x} u_{x y}^{t / b}+\partial_{y}\left(u_{x x}^{t / b}-u_{y y}^{t / b}\right)\right],
\end{equation}
where the piezocoefficient $e_{11}^t = e_{11}^b$ for P-bilayers and $e_{11}^b=-e_{11}^t>0$ for AP-bilayers. 
Computation of piezopotential $\phi^{t/b}$ is determined by taking into account screening-induced charges as well. 
The procedure for such a computation is the one used in Ref. \cite{PhysRevLett.124.206101}, where we considered TMDs to be encapsulated by hexagonal boron nitride.

\section{DFT computations}

Ab initio density-functional-theory calculations were carried out using the Quantum ESPRESSO package\cite{QEcite}. 
We used full relativistic pseudopotentials with spin-orbit interaction included.
The exchange-correlation functional used was the generalized gradient approximation of Perdew-Burke-Ernzerhof (GGA-PBE)\cite{PBE}.
A plane-wave cutoff energy of 1090 eV was used for all calculations. 
The integration over the Brillouin-zone (BZ) was performed
using the scheme proposed by Monkhorst-Pack with a grid of $12 \times 12 \times 1$ $k$-points\cite{MHgrid}.
Monolayer structure parameters of bulk TMDs taken from Refs. \onlinecite{schutte1987crystal} and \onlinecite{bronsema1986structure}
were used for TMD homobilayers. The interlayer hybridization models set out below were parametrized from DFT bands calculated using a range of in-plane relative shifts of the layers, $\boldsymbol{r}_{0}$, and interlayer distances, $d$ (see Fig. \ref{fig:stackings_supl} and Tables \ref{tab:in_plane_shifts} and \ref{tab:interlayer_distances}).

\begin{table}[]
\caption{In-plane relative shifts of the layers used to parametrize the interlayer hybridization models from DFT bands. 
In-plane relative shifts from 1 to 12 were used to extract energies from DFT bandstructure calculations at $\Gamma$- and K-valleys for AP-bilayers. 
In-plane relative shifts from 1 to 7 and 13 to 14 were used to extract energies from DFT bandstructure calculations at both $\Gamma$- and K-valleys for P-bilayers. 
In-plane relative shifts 1, 10, 15 to 24 were used to extract energies from DFT bandstructure calculations at Q-valley  for AP-bilayers. 
In-plane relative shifts 1, 7, 10, 15, 16 and 18 were used to extract energies from DFT bandstructure calculations at Q-valley  for P-bilayer.}
\label{tab:in_plane_shifts}
\begin{tabular}{c|cccc}
\# & $\boldsymbol{r}_{0}/a$ & $H_\Gamma$ & $H_K$ & $H_Q$ \\ \hline
1 (XX$'$)  & $\left(0,0\right)$                                  & AP, P           & AP, P      & AP, P      \\
2  & $\left(0, \frac{1}{4\sqrt{3}} \right)$                &  AP, P           & AP, P &   ---    \\
3  & $\left(0, \frac{1}{3\sqrt{3}} \right)$                &  AP, P           & AP, P &   ---    \\
4  & $\left(0, \frac{1}{2\sqrt{3}} \right)$                &  AP, P           & AP, P &   ---    \\
5 (MM$'$/XM$'$)  & $\left(0,  \frac{1}{\sqrt{3}} \right)$                &  AP, P           & AP, P &   ---    \\
6  & $\left(\frac{1}{8},  \frac{7 \sqrt{3}} {24 }\right)$                &  AP, P           & AP, P &   ---    \\
7 (DW)  & $\left(\frac{1}{4},  \frac{\sqrt{3}} {4 }\right)$                &  AP, P           & AP, P &   P    \\
8  & $\left(\frac{3}{8},  \frac{5 \sqrt{3}} {24 }\right)$                &  AP           & AP &   ---    \\
9 (2H) & $\left(\frac{1}{2},  \frac{ \sqrt{3}} {6 }\right)$               &  AP           & AP &   ---    \\
10  & $\left(\frac{3}{8},  \frac{ \sqrt{3}} {8 }\right)$               &  AP           & AP &   AP, P    \\
11  & $\left(\frac{1}{4},  \frac{\sqrt{3}} {12 }\right)$                 &  AP           & AP &   ---    \\
12  & $\left(\frac{1}{8},  \frac{ \sqrt{3}} {24 }\right)$                &  AP           & AP &   ---    \\
13  & $\left(\frac{1}{6},  \frac{ \sqrt{3}} {6 }\right)$                   & P           & P &   ---    \\
14  & $\left(\frac{1}{12},  \frac{ \sqrt{3}} {8 }\right)$                  & P           & P &   ---    \\ 
15  & $\left(0,  \frac{ \sqrt{3}} {4 }\right)$                      & ---           & --- &   AP, P    \\ 
16  & $\left(\frac{1}{4}, 0\right)$         
& ---           & --- &   AP, P    \\
17  & $\left(-\frac{1}{4}, 0\right)$              
 & ---           & --- &   AP    \\
18  & $\left(-\frac{1}{8},  \frac{ \sqrt{3}} {8 }\right)$           & ---           & --- &   AP, P    \\
19  & $\left(\frac{1}{8},  -\frac{ \sqrt{3}} {8 }\right)$           & ---           & --- &   AP    \\
20  & $\left(\frac{1}{8},  \frac{ \sqrt{3}} {8 }\right)$            & ---           & --- &   AP    \\
21  & $\left(-\frac{1}{8},  -\frac{ \sqrt{3}} {8 }\right)$          & ---          & --- &   AP    \\
22  & $\left(0, \frac{ 1}    {\sqrt{3} }\right)$            
 & ---          & --- &   AP    \\
23  & $\left(\frac{1}{4}, \frac{ \sqrt{3}} {4}\right)$              & ---           & --- &   AP    \\    
24  & $\left(\frac{1}{2}, \frac{ 1} {2 \sqrt{3}}\right)$            & ---           & --- &   AP    \\   \hline
\end{tabular}
\end{table}

\begin{table}[]
\caption{Interval of interlayer distances that were used to compute DFT bandstructure calculations using in-plane relative shifts  listed in Table \ref{tab:in_plane_shifts}.}
\label{tab:interlayer_distances}
\begin{tabular}{c|ccccc}
\# & $d$ \AA & MoS$_2$          & MoSe$_2$ & WS$_2$ & WSe$_2$ \\ \hline
1 & 6.0      &  $\Gamma$, K, Q  &  ---            &   $\Gamma$, K, Q    &   ---\\
2  & 6.1     &  $\Gamma$, K     &  ---            &   $\Gamma$, K &  ---\\
3  & 6.1489  &  $\Gamma$, K, Q  &  ---            &   --- &  ---\\
4  & 6.1725  &  --- &  ---      &   $\Gamma$, K, Q  & --- \\
5  & 6.2     &  $\Gamma$, K     &  ---            &   $\Gamma$, K &  --- \\
6  & 6.3     &  $\Gamma$, K, Q  &  $\Gamma$, K           &  $\Gamma$, K, Q &    $\Gamma$, K  \\
7  & 6.4     &  $\Gamma$, K     &  $\Gamma$, K, Q            &   $\Gamma$, K  &  $\Gamma$, K, Q \\
8  & 6.463   &  ---             &  $\Gamma$, K, Q            &   & ---   \\
9  & 6.477   &  ---             &  ---            &   ---  & $\Gamma$, K, Q  \\
10  & 6.5     &  Q               &  $\Gamma$, K        &   Q & $\Gamma$, K \\
11  & 6.6     & $\Gamma$, K      &  $\Gamma$, K, Q            &  $\Gamma$, K  & $\Gamma$, K, Q \\
12  & 6.7     &  $\Gamma$, K, Q  &  $\Gamma$, K           &   $\Gamma$, K, Q  &  $\Gamma$, K  \\
13  & 6.8     &  $\Gamma$, K     &  $\Gamma$, K, Q            &   $\Gamma$, K &  $\Gamma$, K, Q \\
14  & 6.9     &  ---             &  $\Gamma$, K          &   ---   & $\Gamma$, K\\
15 & 7.0     & $\Gamma$, K      &  $\Gamma$, K, Q            &  $\Gamma$, K   &  $\Gamma$, K, Q    \\
16 & 7.2     & $\Gamma$, K      &  $\Gamma$, K            &   $\Gamma$, K &  $\Gamma$, K \\
17  & 7.4    & ---              &  $\Gamma$, K           &   ---  & $\Gamma$, K\\ \hline
\end{tabular}
\end{table}


\section{Interlayer hybridization models}

To determine modulation of valence band edges at $\Gamma$, K-valleys and conduction band edge at K- and Q-valleys in twisted TMD homobilayers, we, first, establish effective Hamiltonians describing the interlayer coupling of the corresponding states in aligned AP- and P-bilayers for a given lateral offset $\bm{r}_0$ and interlayer distance $d$  between the layers (see Fig. \ref{fig:stackings_supl}). In these effective Hamiltonians we use translational symmetry of the aligned bilayers and represent all matrix elements in terms of Fourier series over reciprocal vectors of a TMD monolayer, keeping only the lowest harmonics and accounting for symmetry of restrictions.

For $\Gamma$-valley, an effective Hamiltonian can be represented as: $H_{\Gamma,\rm VB}^{P / A P} = H_{\Gamma} +\delta H_{\Gamma}^{P}+\delta H_{\Gamma}^{AP}$, with
\begin{equation}
\label{eq:ef_H_AP_and_P_VB_GAMMA}
H_{\Gamma}  = 
\begin{bmatrix}
 \varepsilon_{\Gamma}\left(\boldsymbol{r}_{0}, d\right) & T_{\Gamma}\left(\boldsymbol{r}_{0}, d\right) \\
T_{\Gamma}\left(\boldsymbol{r}_{0}, d\right)&\varepsilon_{\Gamma}\left(\boldsymbol{r}_{0}, d\right)
\end{bmatrix},
\end{equation}
\begin{equation}\label{eq:interlayer_charge_transfer}
\delta H_{\Gamma}^{P}  = 
\begin{bmatrix}
-\frac{\Delta^{P}\left(\boldsymbol{r}_{0}, d\right)}{2} &  0 \\
 0 & \frac{\Delta^{P}\left(\boldsymbol{r}_{0}, d\right)}{2} 
\end{bmatrix},
\end{equation}
and
\begin{equation}\label{eq:tunneling_amendment}
\delta H_{\Gamma}^{AP}  = 
\begin{bmatrix}
 0 &  \delta T_{\Gamma}^{AP}\left(\boldsymbol{r}_{0}, d\right) \\
 \delta T_{\Gamma}^{AP}\left(\boldsymbol{r}_{0}, d\right) & 0 
\end{bmatrix}.
\end{equation}
Hamiltonians (\ref{eq:ef_H_AP_and_P_VB_GAMMA}-\ref{eq:tunneling_amendment}) act on a two-component wave function, where the first component describes the top layer state, the second -- states of the bottom layer. The first term \eqref{eq:ef_H_AP_and_P_VB_GAMMA}, responsible for the interlayer hybridization of the monolayer states, is the same for P- and AP-bilayers. The second one \eqref{eq:interlayer_charge_transfer}, describing the electron potential energy shift, $\Delta^P$, induced by the interlayer charge transfer, is applied only for P-bilayers. The last term \eqref{eq:tunneling_amendment} takes into account $\bm{r}_0\to-\bm{r}_0$ asymmetry of resonant coupling in AP-bilayers. Matrix elements in Eqs. (\ref{eq:ef_H_AP_and_P_VB_GAMMA}-\ref{eq:tunneling_amendment}) read as follows:
\begin{equation}\label{eq:matrix_elements_H_PandAP}
\begin{split}
\varepsilon_{\Gamma}(\bm{r}_0,d)& = \varepsilon_{A'} +  v_{\Gamma,0}(d) + v_{\Gamma,1}(d)\sum_{j=1,2,3}\cos(\bm{G}_j\cdot\bm{r_0}),\\
T_{\Gamma}(\bm{r}_0,d) & =  \frac{t_0(d)}{2}+\frac{t_1(d)}{2}\sum_{j=1,2,3}\cos(\bm{G}_j\cdot\bm{r_0}), \\
\Delta^P(\bm{r}_0,d) & = \Delta_a(d)\sum_{j=1,2,3}\sin(\bm{G}_j\cdot\bm{r_0}), \\
\delta T_{\Gamma}^{AP}(\bm{r}_0,d) &= \frac{t_2^{AP}(d)}{2}\sum_{j=1,2,3}\sin(\bm{G}_j\bm{r_0}).
\end{split}
\end{equation}
The interlayer distance dependence of the matrix elements \eqref{eq:matrix_elements_H_PandAP} were extracted from fitting of the $\Gamma$-valley valence band edge in P- and AP- TMD bilayers calculated in DFT for various offsets and interlayer distances listed in Tables \ref{tab:in_plane_shifts} and \ref{tab:interlayer_distances}. Interlayer distances are chosen by taking the $\boldsymbol{r}_{0}$ dependence found using the methods set out in Ref. \onlinecite{PhysRevLett.124.206101}, then rigidly shifting the resulting values of $d$ such that the value for 2H stacking agrees with the experimentally determined quantity for bulk crystals. Results of the fitting are gathered in Table \ref{tab:fittings_GAMMA}.
\begin{table}[h]
\caption{Fitting parameters for $d$-dependent functions in Eq. (\ref{eq:matrix_elements_H_PandAP}) using the expression $Ae^{-q(d-d'_0)}$, where  $d'_0$ result from a shift applied to $d_0$ (Table \ref{tab:fittings_GAMMA}) so that d$_{2H}$ obtained from DFT calculation matches the experimental values taken from  
Refs. \onlinecite{schutte1987crystal} and \onlinecite{bronsema1986structure}.}
    \begin{tabular}{lllllllll}
    \label{tab:fittings_GAMMA}
    & \multicolumn{2}{c}{MoS$_2$} & \multicolumn{2}{c}{MoSe$_2$} & \multicolumn{2}{c}{WS$_2$} & \multicolumn{2}{c}{WSe$_2$} \\\hline \hline
    & $A, \mathrm{eV} \quad $       & $q$, \AA$^{-1} \qquad$         & $A, \mathrm{eV} \qquad $       & $q$, \AA$^{-1} \qquad$ & $A, \mathrm{eV} \quad $       & $q$, \AA$^{-1} \quad$   &$A, \mathrm{eV} \qquad $       & $q$, \AA$^{-1} \qquad$   \\ \hline \hline
    $t_0\qquad $                &  0.638   &   1.085  &  0.598   &   1.084 &  0.615   &   1.005    &   0.578   &   1.007        \\
    $t_1\qquad $                &  0.029   &   2.545  &  0.022   &   2.518 &  0.026   &   2.488    &   0.021   &   2.499        \\
    $t_2^{A P}\qquad  $         &  0.003   &   2.288  &  0.003   &   2.196 &  0.004   &   2.400    &   0.003   &   2.334        \\   
    $v_{\Gamma,0}\qquad $       &  0.137   &   1.856  &  0.141   &   1.770 &  0.116   &   1.810    &   0.123   &   1.732        \\  
    $v_{\Gamma,1}\qquad  $      &  0.006   &   3.221  &  0.005   &   3.104 &  0.004   &   3.314    &   0.004   &   3.219        \\  
    $\Delta_a\qquad $           &  0.016   &   2.215  &  0.016   &   2.052 &  0.015   &   2.260    &   0.015   &   2.095        \\    
    $\varepsilon_{A'}\qquad$    &  -6.074  &    ---   &  -5.627  &   ---   &  -5.824  &   ---      &   -5.393   &   ---         \\      
    \hline \hline
    $d'_0$,\AA  & \multicolumn{2}{c}{6.36} & \multicolumn{2}{c}{6.70} & \multicolumn{2}{c}{6.38} & \multicolumn{2}{c}{6.71}     \\ \hline \hline
    \end{tabular}
\end{table}

Effective Hamiltonians describing coupling of the valence band at the K-valley for P- and AP-bilayers are given by
\begin{equation}
\label{eq:ef_H_P_VB_K}
H_{\rm VB,K}^{P,\tau}=\begin{bmatrix}
\varepsilon_{\rm VB,K}^{P}\left(\bm{r}_{0},d\right)-\frac{\Delta^{P}\left(\bm{r}_{0},d\right)}{2} & T_{\rm VB,K}^{P}\left(\bm{r}_{0},d  \right) \\
T_{\rm VB,K}^{P *}\left(\bm{r}_{0}, d\right) & \varepsilon_{\rm VB,K}^{P}\left(\bm{r}_{0},d\right)+\frac{\Delta^{P}\left(\bm{r}_{0},d\right)}{2}
\end{bmatrix},
\end{equation}
\begin{equation}\label{eq:ef_H_AP_VB_K}
	H_{\rm VB,K}^{AP,\tau}=\begin{bmatrix}
	\varepsilon_{\rm VB,K}^{AP}(\bm{r}_0,d) +\tau s \frac{\Delta_{\rm VB}^{SO}(\bm{r}_0,d)}{2}   & T_{\mathrm{VB,K}}^{AP}(\bm{r}_0,d)\\ \\
	T_{\rm VB,K}^{AP *}(\bm{r}_0,d) &  \varepsilon_{\rm VB,K}^{AP}(\bm{r}_0,d) - \tau s \frac{\Delta_{\rm VB}^{SO}(\bm{r}_0,d)}{2}
	\end{bmatrix}
\end{equation}
where the matrix elements read as follows:
\begin{equation}\label{eq:H_VB_PandAP_terms}
\begin{split}
	T_{\rm VB,K}^P(\bm{r}_0,d) =& t_{\rm VB}^P\left(d\right)\sum_{j=0,1,2}e^{iC_3^j \tau\mathbf{K}\cdot\bm{r}_0},\\
	T_{\rm VB,K}^{AP}(\bm{r}_0,d) =& t_{{\rm VB}}^{AP}\left(d\right)\sum_{j=0,1,2}e^{iC_3^j \tau\mathbf{K}\cdot\bm{r}_0}e^{i\tfrac{2\pi}{3}\tau j},\\
	\varepsilon_{\rm VB,K}^P(\bm{r}_0,d) =& \varepsilon_{\rm VB'}^{P} - v_{{\rm VB},0}^P\left(d\right) - v_{{\rm VB},1}^{P}\left(d\right)\sum_{j=1,2,3}\cos{\left(\mathbf{G}_j\cdot\bm{r}_0 \right)},\\
	\varepsilon_{\rm VB,K}^{AP}(\bm{r}_0,d) = &  \varepsilon_{{\rm VB'}}^{AP} - v_{{\rm VB},0}^{AP}\left(d\right) - v_{{\rm VB},1}^{AP}\left(d\right)\sum_{j=1,2,3}\cos{\left(\mathbf{G}_j\cdot\bm{r}_0\right)}- v_{{\rm VB},2}^{AP}\left(d\right)\sum_{j=1,2,3}\sin{\left(\mathbf{G}_j\cdot\bm{r}_0\right)},\\
	\Delta_{{\rm VB}}^{SO}(\bm{r}_0,d) =&  \Delta_{SO,\rm VB}^{AP}+\Delta_{\rm VB,0}^{AP}\left(d\right)+\Delta_{\rm VB,1}^{AP}\left(d\right)\sum_{j=1,2,3}\cos{\left(\mathbf{G}_j\cdot\bm{r}_0 \right)}+\Delta_{\rm VB,2}^{AP}\left(d\right)\sum_{j=1,2,3}\sin{\left(\mathbf{G}_j\cdot\bm{r}_0 \right)}.
\end{split}
\end{equation}
The quantity $\Delta_{{\rm VB}}^{SO} $ is the spin-orbit splitting in the correspond band, $\tau$ and $s=-\tau$ is the valley  and spin index respectively. 
The other terms were already defined above.
The  effective hybridization Hamiltonian for conduction band at the K-valley for P and AP are given by

\begin{equation}\label{eq:ef_H_P_CB_K}
    H_{\rm CB,K}^{P,\tau}=\begin{bmatrix}
    \varepsilon_{\rm CB,K}^{P}(\bm{r}_0,d)-\tfrac{\Delta^P(\bm{r}_0,d)}{2} & T_{\rm CB,K}^P(\bm{r}_0,d)\\
    T_{\rm CB,K}^{P*}(\bm{r}_0,d) & \varepsilon_{\rm CB,K}^P(\bm{r}_0,d) + \tfrac{\Delta^P(\bm{r}_0,d)}{2}
    \end{bmatrix},
\end{equation}

\begin{equation}\label{eq:ef_H_AP_CB_K}
    H_{\rm CB,K}^{AP,\tau}=\begin{bmatrix}
    \varepsilon_{\rm CB,K}^{AP}(\bm{r}_0,d) - \tfrac{\tau s \Delta_{\rm CB}^{SO}(\bm{r}_0,d)}{2} & T_{\rm CB,K}^{AP}(\bm{r}_0,d)\\
    T_{\rm CB,K}^{AP*}(\bm{r}_0,d) & \varepsilon_{\rm CB,K}^{AP}(\bm{r}_0,d) +  \tfrac{\tau s\Delta_{\rm CB,K}^{SO}(\bm{r}_0,d)}{2}
    \end{bmatrix},
\end{equation}
where the matrix elements read
\begin{equation}\label{eq:H_CB_PandAP_terms}
\begin{split}
	\varepsilon_{\rm CB,K}^{P}(\bm{r}_0,d) = &  \varepsilon_{{\rm CB'}}^{P} - v_{{\rm CB},0}^{P}\left(d\right) - v_{{\rm CB},1}^{P}\left(d\right)\sum_{j=1,2,3}\cos{\left(\mathbf{G}_j\cdot\bm{r}_0\right)},\\
	\varepsilon_{\rm CB,K}^{AP}(\bm{r}_0,d) = &  \varepsilon_{{\rm CB'}}^{AP} - v_{{\rm CB},0}^{AP}\left(d\right) - v_{{\rm CB},1}^{AP}\left(d\right)\sum_{j=1,2,3}\cos{\left(\mathbf{G}_j\cdot\bm{r}_0\right)} - v_{{\rm CB},2}^{AP}\left(d\right)\sum_{j=1,2,3}\sin{\left(\mathbf{G}_j\cdot\bm{r}_0\right)},\\
	\Delta_{{\rm CB}}^{SO}(\bm{r}_0,d) =&  \Delta_{SO,\rm CB}^{AP}+\Delta_{\rm CB,1}^{AP}\left(d\right)\sum_{j=1,2,3}\cos{\left(\mathbf{G}_j\cdot\bm{r}_0 \right)}+\Delta_{\rm CB,2}^{AP}\left(d\right)\sum_{j=1,2,3}\sin{\left(\mathbf{G}_j\cdot\bm{r}_0 \right)},\\
	 T_{\rm CB,K}^{P/AP}(\bm{r}_0,d)&=t_{\rm CB}^{P/AP}\left(d\right)\sum_{j=0,1,2}e^{i\tau C_3^j \mathbf{K}\cdot \bm{r}_0}.
\end{split}
\end{equation}
Fitting parameters for quantities on the left hand side of Eqs. (\ref{eq:H_VB_PandAP_terms}-\ref{eq:H_CB_PandAP_terms}) using the expression $Ae^{-q(d-d'_0)}$ are shown in Table \ref{tab:fittings_K}.
\begin{table}[h]
\caption{Fitting parameters for quantities on left hand side of Eqs. (\ref{eq:H_VB_PandAP_terms}-\ref{eq:H_CB_PandAP_terms}) using the expression $Ae^{-q(d-d'_0)}$ where values for $d'_0$ are listed in Table \ref{tab:fittings_GAMMA}.}
    \begin{tabular}{lllllllll}
       & \multicolumn{2}{c}{MoS$_2$} & \multicolumn{2}{c}{MoSe$_2$} & \multicolumn{2}{c}{WS$_2$} & \multicolumn{2}{c}{WSe$_2$} \\\hline \hline
       & $A, \mathrm{eV} \quad $       & $q$, \AA$^{-1} \qquad$         & $A, \mathrm{eV} \qquad $       & $q$, \AA$^{-1} \qquad$ & $A, \mathrm{eV} \quad $       & $q$, \AA$^{-1} \quad$   &$A, \mathrm{eV} \qquad $       & $q$, \AA$^{-1} \qquad$   \\ \hline \hline
    $v_{\rm VB,0}^{P}\qquad$            &    0.008   &   1.746       &   0.009   &   1.715       &   0.009    &   1.84     &    0.011   &   1.809      \\
    $v_{\rm VB,1}^{P}\qquad$            &    0.002   &   3.049       &   0.002   &   2.826       &   0.002    &   3.023    &    0.003   &   2.783      \\
    $v_{\rm VB,0}^{AP}\qquad$           &    0.007   &   1.869       &   0.008   &   1.704       &   0.007    &   1.976    &    0.010   &   1.726       \\
    $v_{\rm VB,1}^{AP}\qquad$           &    0.001   &   3.104       &   0.001   &   2.837       &   0.001    &   2.912    &    0.001   &   2.864      \\
    $v_{\rm VB,2}^{AP}\qquad$           &   -0.001   &   3.412       &  -0.001   &   3.316       &   -0.001   &   3.341    &    -0.001  &   3.127     \\
    $v_{\rm CB,0}^{P}\qquad$            &    0.007   &   2.199       &   0.007   &   2.133       &   0.007    &   2.137    &    0.008   &   2.033      \\
    $v_{\rm CB,1}^{P}\qquad$            &    0.001   &   2.903       &   0.001   &   2.943       &   0.001    &   3.105    &    0.002   &   2.945      \\
    $v_{\rm CB,0}^{AP}\qquad$           &    0.006   &   2.308       &   0.007   &   2.097       &   0.006    &   2.411    &    0.007   &   2.125      \\
    $v_{\rm CB,1}^{AP}\qquad$           &    0.001   &   3.064       &   0.001   &   2.996       &   0.001    &   2.941    &    0.001   &   3.130       \\
    $v_{\rm CB,2}^{AP}\qquad$           &    0.001   &   2.815       &   0.001   &   2.574       &   0.000    &   2.048    &    0.000   &   2.333        \\
    $|t_{\rm VB}^{P}|\qquad$            &    0.012   &   1.616       &   0.013   &   1.517       &   0.014    &   1.600    &    0.018   &   1.495      \\
    $|t_{\rm VB}^{AP}|\qquad$           &   -0.011   &   1.589       &   0.015   &   1.889       &   -0.000   &   4.657    &    0.005   &   0.000        \\
    $|t_{\rm CB}^{P}|\qquad$            &    0.002   &   1.779       &   0.003   &   1.473       &   0.003    &   1.466    &    0.004   &   1.425      \\
    $|t_{\rm CB}^{AP}|\qquad$           &   -0.000   &   1.652       &   0.002   &   1.133       &   0.003    &   1.096    &    0.004   &   1.040       \\
    $\Delta_{\rm VB,0}^{AP}\qquad$      &   -0.033   &   0.005       &   -0.003  &   4.785       &   0.001    &   3.650    &    0.001   &   3.134      \\
    $\Delta_{\rm VB,1}^{AP}\qquad$      &   -0.000   &   2.841       &   0.001   &   5.191       &   -0.001   &   3.079    &    -0.001  &   3.046     \\
    $\Delta_{\rm VB,2}^{AP}\qquad$      &   -0.000   &   3.506       &   0.002   &   5.022       &   -0.001   &   3.310    &    -0.001  &   3.076     \\
    $\Delta_{\rm CB,1}^{AP}\qquad$      &   -0.002   &   1.946       &   -0.002  &   2.666       &   -0.001   &   2.350    &    -0.002  &   2.139     \\
    $\Delta_{\rm CB,2}^{AP}\qquad$      &    0.003   &   2.164       &   0.002   &   2.521       &   -0.000   &   4.431    &    0.000   &   0.000       \\
    $\varepsilon_{\rm VB'}^{P}\qquad$    &  -5.844   &    ---        &  -5.203   &     ---       &  -5.495    &    ---     &   -4.836   &      ---    \\
    $\varepsilon_{\rm VB'}^{AP}\qquad$   &  -5.905   &    ---        &  -5.31    &     ---       &  -5.698    &    ---     &   -5.070   &      ---     \\
    $\varepsilon_{\rm CB'}^{P}\qquad$    &  -4.076   &    ---        &  -3.736   &     ---       &  -3.808    &    ---     &   -3.462   &     ---     \\
    $\varepsilon_{\rm CB'}^{AP}\qquad$   &  -4.061   &    ---        &  -3.735   &     ---       &   -3.777   &    ---     &    -3.44   &      ---    \\
    $\Delta_{SO,\rm VB}^{AP}\qquad$     &    0.183   &    ---        &   0.19    &     ---       &   0.432    &     ---    &    0.468   &      ---   \\
    $\Delta_{SO,\rm CB}^{AP}\qquad$     &    0.005   &    ---        &   0.023   &     ---       &   0.037    &     ---    &    0.044   &      ---    \\ \hline \hline
    \end{tabular}
\label{tab:fittings_K}
\end{table}

For Q$_1$-valley (see Fig. \ref{fig:BZ}), the effective Hamiltonian is expressed as:
\begin{equation}\label{eq:ef_H_P_CB_Q}
    H_{\text{CB},\mathrm{Q}_1}^{P/AP}=\begin{bmatrix}
    \varepsilon_{\rm CB,\mathrm{Q}_1}^{P/AP}\left(\bm{r}_0,d\right)-\frac{S^{P / AP}\left(\bm{r}_0,d\right)}{2} & T_{\rm Q_{1}}^{P / AP}\left(\bm{r}_0,d\right)\\
    T_{\rm Q_{1}}^{P/AP*} \left(\bm{r}_0,d\right)& \varepsilon_{\rm CB,\mathrm{Q}_1}^{P/AP}\left(\bm{r}_0,d\right)+\frac{S^{P / AP}\left(\bm{r}_0,d\right)}{2}
    \end{bmatrix},
\end{equation}
where the matrix elements  are given by
\begin{equation}
\label{eq:H_CB_Q_PandAP_terms}
\begin{split}
\varepsilon_{\rm CB,\mathrm{Q}_1}^{AP}\left(\bm{r}_0,d\right) &=\epsilon_{\rm Q}^{AP}+v_{\rm Q,0}^{AP}\left(d\right)+\sum_{j=1,2,3}\left[v_{\rm Q,j}^{AP,s}\left(d\right) \cos \left(\boldsymbol{G}_{j} \boldsymbol{r}_{0}\right)+v_{\rm Q,j}^{AP,a}\left(d\right) \sin \left(\boldsymbol{G}_{j} \boldsymbol{r}_{0}\right)\right], \\
\varepsilon_{\rm CB,\mathrm{Q}_1}^{P}\left(\bm{r}_0,d\right) &=\varepsilon_{\rm Q}^P+v_{\rm Q,0}^P\left(d\right)+\sum_{j=1,2,3} v_{\rm Q,j}^{P,s}\left(d\right) \cos \left(\boldsymbol{G}_{j} \boldsymbol{r}_{0}\right), \\
T_{\rm Q_{1}}^{A P}\left(\bm{r}_0,d\right) &=\left|t_{\rm Q,0}\right|^{AP}\left(d\right) +\left|t_{\rm Q,1}\right|^{AP} \left(d\right) e^{-i \boldsymbol{G}_{1} \boldsymbol{r}_{0}+i \varphi_{10}}+\left|t_{\rm Q,2}\right|^{AP}\left(d\right)  e^{i \boldsymbol{G}_{2} \boldsymbol{r}_{0}+i \varphi_{20}}+\left|t_{\rm Q,3+}\right|^{AP}\left(d\right)  e^{i \boldsymbol{G}_{3} \boldsymbol{r}_{0}+i \varphi_{30+}}+\left|t_{\rm Q,3-}\right|^{AP}\left(d\right) e^{-i \boldsymbol{G}_{3} \boldsymbol{r}_{0}+i \varphi_{30}-}, \\
T_{\rm Q_{1}}^{P}\left(\bm{r}_0,d\right) &=\left|t_{\rm Q,0}\right|^{P}\left(d\right)+\left|t_{\rm Q,1}\right|^{P}\left(d\right) e^{-i \boldsymbol{G}_{1} \boldsymbol{r}_{0}}+\left|t_{\rm Q,2}\right|^{P}\left(d\right) e^{i \boldsymbol{G}_{2} \boldsymbol{r}_{0}}+\left|t_{\rm Q,3+}\right|^{P}\left(d\right) e^{i \boldsymbol{G}_{3} \boldsymbol{r}_{0}}+\left|t_{\rm Q,3-}\right|^{P}\left(d\right) e^{-i \boldsymbol{G}_{3} \boldsymbol{r}_{0}}, \\
S^{P}\left(\bm{r}_0,d\right) &=\Delta_{a}^{\rm Q} \left(d\right)\sum_{j=1,2,3} \sin \left(\boldsymbol{G}_{j} \boldsymbol{r}_{0}\right),\\
S^{A P} &=\Delta_{SO}^{\rm Q},
\end{split}
\end{equation}
where $\Delta_{SO}^{\rm Q}$ is the monolayer spin-orbit splitting at Q. 
In Table \ref{tab:fittings_Q} are listed the fitting parameters for the left hand side quantities in Eqs. \ref{eq:H_CB_Q_PandAP_terms}.
Matrix elements for Q$_{2}$ and Q$_{3}$ can be obtained by applying the $C_3^{1}$ and $C_3^{-1}$ rotation operations, respectively, to $G_{1,2,3}$ in phase factors of terms in Eqs. \ref{eq:H_CB_Q_PandAP_terms}. 
Energies of $-$Q$_{1,2,3}$ correspond to those of Q$_{1,2,3}$  respectively. 
\begin{table}[h]
\caption{Fitting parameters for quantities on left hand side of Eqs. \ref{eq:H_CB_Q_PandAP_terms} using the expression $Ae^{-q(d-d_0)}$ where values for $d_0$ are listed in Table \ref{tab:fittings_GAMMA}.}
    \begin{tabular}{lllllllll}
       & \multicolumn{2}{c}{MoS$_2$} & \multicolumn{2}{c}{MoSe$_2$} & \multicolumn{2}{c}{WS$_2$} & \multicolumn{2}{c}{WSe$_2$} \\\hline \hline
       & $A, \mathrm{eV} \quad $       & $q$, \AA$^{-1} \qquad$         & $A, \mathrm{eV} \qquad $       & $q$, \AA$^{-1} \qquad$ & $A, \mathrm{eV} \quad $       & $q$, \AA$^{-1} \quad$   &$A, \mathrm{eV} \qquad $       & $q$, \AA$^{-1} \qquad$   \\ \hline \hline
    $\left|t_{\rm Q,0}\right|^{AP}\qquad$                    &     0.169     & 0.694     &     0.176 & 0.649       &     0.168 & 0.672   &     0.188 & 0.591     \\
    $\left|t_{\rm Q,1}\right|^{AP}\qquad$                    &     0.005     & 1.607     &     0.005 & 1.637       &     0.008 & 1.662   &     0.007 & 1.808     \\
    $\left|t_{\rm Q,2}\right|^{AP}\qquad$                    &    0.004      & 1.69      &    0.005  & 1.706       &    0.007 & 1.806    &    0.007  & 1.796    \\
    $\left|t_{\rm Q,3\pm}\right|^{AP}\qquad$                    &    0.003      & 2.233     &    0.002  & 1.931       &    0.004 & 2.251    &    0.002  & 1.894   \\
    $\left|t_{\rm Q,0}\right|^{P}\qquad$                   &    0.164      & 0.711     &    0.176  & 0.672       &    0.154 & 0.709    &    0.169  & 0.669   \\
    $\left|t_{\rm Q,1}\right|^{P}\qquad$                   &    0.011      & 2.494     &    0.011  & 2.358       &    0.012 & 2.327    &    0.012  & 2.21   \\
    $\left|t_{\rm Q,2}\right|^{P}\qquad$                   &    0.011      & 2.494     &    0.011  & 2.358       &    0.012 & 2.327    &    0.012  & 2.21    \\
    $\left|t_{\rm Q,3\pm}\right|^{P}\qquad$                   &    0.004      & 1.978     &    0.002  & 1.861       &    0.003 & 2.126    &    0.003  & 1.93   \\
    $v_{\rm Q,1}^{AP,s}\qquad$                 &     -0.005 & 2.247     &     -0.005& 2.378       &   -0.004 & 2.596    &     -0.006 & 2.165    \\
    $v_{\rm Q,2}^{AP,s}\qquad$                 &    -0.006  & 2.593     &    -0.006 & 2.399       &    -0.005 & 2.818   &    -0.006 & 2.407     \\
    $v_{\rm Q,3}^{AP,s}\qquad$                 &    -0.001  & 2.058     &    -0.001 & 3.076       &    0.001 & 2.865    &    -0.001 & 0.135     \\
    $v_{\rm Q,1}^{AP,a}\qquad$                 &    0.005   & 2.866     &    0.004  & 2.89        &    0.006 & 2.822    &    0.003  & 3.565   \\
    $v_{\rm Q,2}^{AP,a}\qquad$                 &    0.005   & 2.817     &    0.004  & 2.89        &    0.006 & 2.684    &    0.003  & 3.565   \\
    $v_{\rm Q,3}^{AP,a}\qquad$                 &    -0.003  & 2.674     &    -0.002 & 2.661       &    -0.003 & 2.516   &    -0.002 & 3.025     \\
    $v_{\rm Q,1}^{P,s}\qquad$                  &    -0.003  & 2.886     &    -0.004 & 2.821       &    -0.003 & 2.753   &    -0.004 & 2.758     \\
    $v_{\rm Q,2}^{P,s}\qquad$                  &    -0.003  & 2.886     &    -0.004 & 2.821       &    -0.003 & 2.753   &    -0.004 & 2.758     \\
    $v_{\rm Q,3}^{P,s}\qquad$                  &    -0.002  & 2.334     &    -0.002 & 2.302       &    -0.002 & 2.154   &    -0.002 & 2.32      \\
    $v_{\rm Q,0}^{AP}\qquad$                  &     -0.007  & 2.666     &    -0.013 & 1.615       &    -0.002 & 3.165   &    -0.003 & 2.851     \\
    $v_{\rm Q,0}^{P}\qquad$                  &      -0.011  & 1.311     &    0.004  & -3.054      &    0.002 & -1.815   &    0.00    & -5.758   \\
    $\Delta_a^Q \qquad$                    &   0.026   &  1.911    &   0.018   &  2.623      &   0.024 &  2.064    &    0.020   &  2.928  \\
    $\varepsilon_{\rm Q}^{AP}\qquad$           &    -4.044  &   ---     &   -3.684  &   ---       &    -3.705 &  ---    &    -3.392 &    ---   \\
    $\varepsilon_{\rm Q}^{P}\qquad$            &   -4.088   &   ---     &   -3.700  &   ---       &   -3.850 &    ---    &    -3.502 &    ---    \\
    $\Delta_{SO}^{\rm Q}  $                       &   0.065    &   ---     &   0.022   &   ---       &   0.257 &    ---    &    0.192  &    ---  \\
    $\phi_{10}\qquad$                      &    -0.54$\pi$   &   ---     &   -0.482$\pi$  &   ---       &   -0.462$\pi$ &   ---    &    -0.444 $\pi$&    ---   \\
    $\phi_{20}\qquad$                      &    0.543$\pi$   &   ---     &    0.481$\pi$  &   ---       &   0.464$\pi$ &    ---    &   0.444 $\pi$  &    --- \\
    $\phi_{30+} = -\phi_{30-} \qquad$       &   0.378$\pi$    &   ---     &    0.372$\pi$  &   ---       &   0.434$\pi$ &    ---    &   0.335 $\pi$  &    ---  \\ \hline \hline
    \end{tabular}
\label{tab:fittings_Q}
\end{table}

\begin{figure}[h]
\includegraphics[scale=0.24]{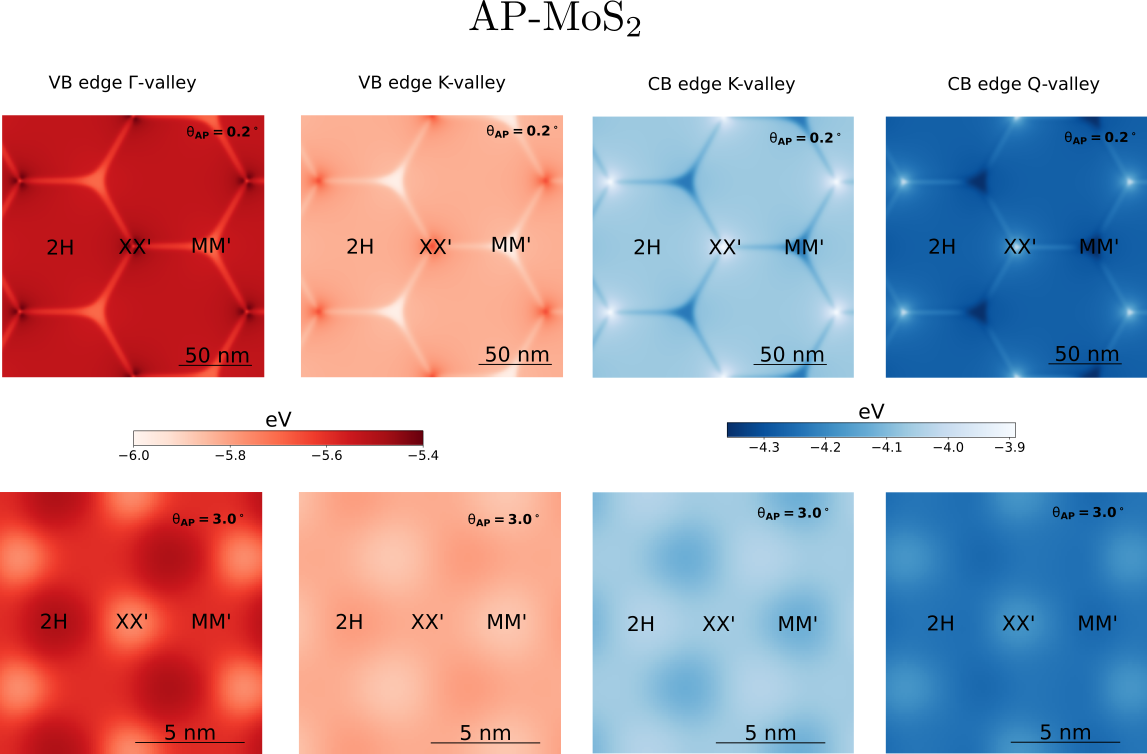}
\caption{Maps of band edge for AP-MoS$_2$ homobilayer in the VB at the $\Gamma$- and K-valley and CB at the K and Q -valley for $\theta_{AP} = 0.2^\circ$ and $\theta_{AP} = 3^\circ$.
See Fig. \ref{fig:stackings_supl} for more details about 2H, XX$'$ and MM$'$ stacking configurations.}
\label{fig:APMoS2_band_maps_supl}
\end{figure}

\begin{figure}[h]
\includegraphics[scale=0.24]{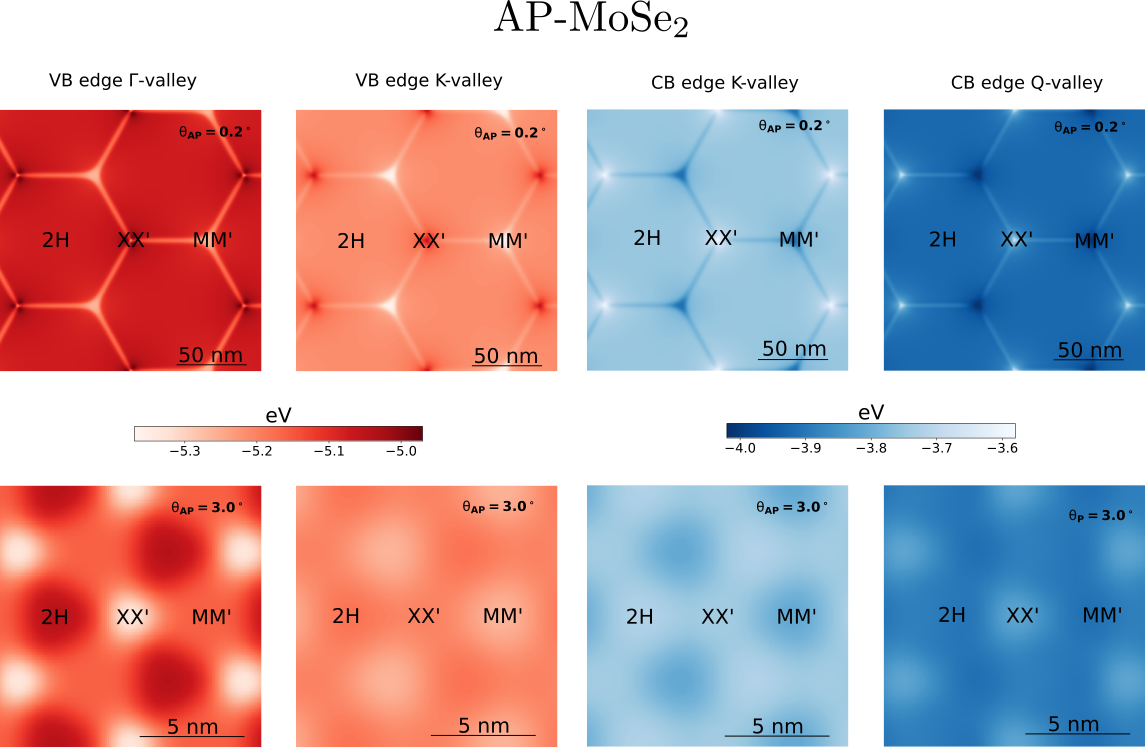}
\caption{Same as in Fig. \ref{fig:APMoS2_band_maps_supl} for AP-MoSe$_2$}
\label{fig:APMoSe2_band_maps_supl}
\end{figure}

\begin{figure}[h]
\includegraphics[scale=0.24]{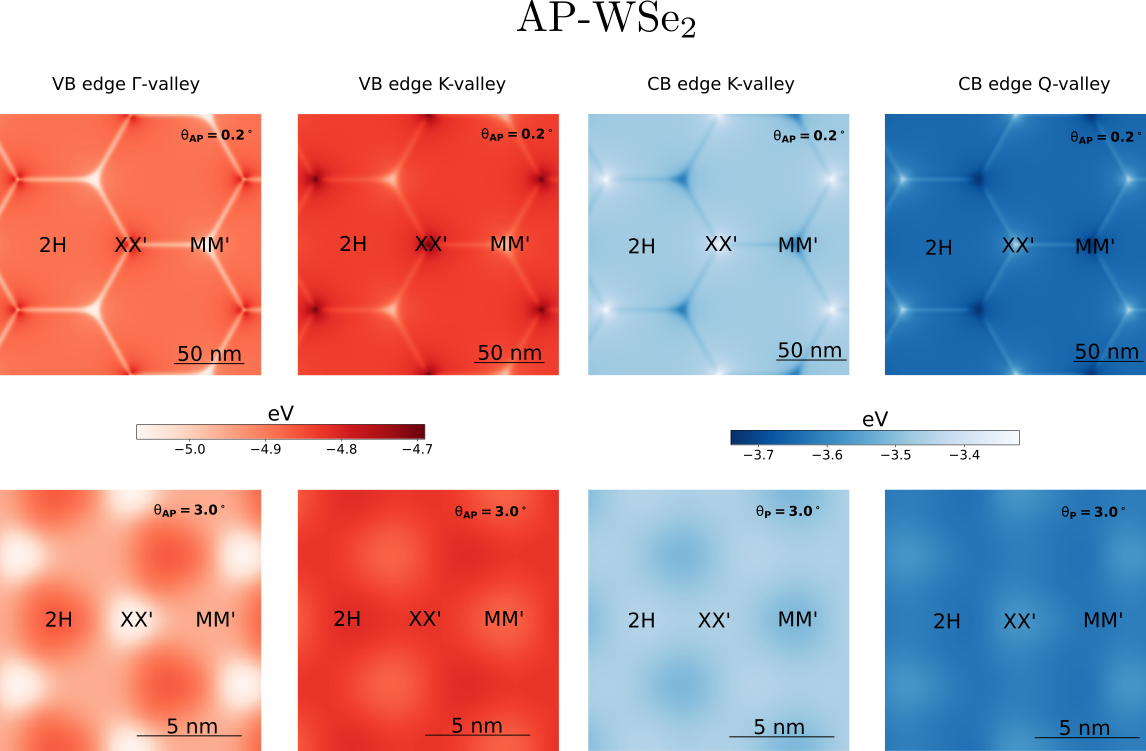}
\caption{Same as in Fig. \ref{fig:APMoS2_band_maps_supl} for AP-WSe$_2$}
\label{fig:APWSe2_band_maps_supl}
\end{figure}

\begin{figure}[h]
\includegraphics[scale=0.24]{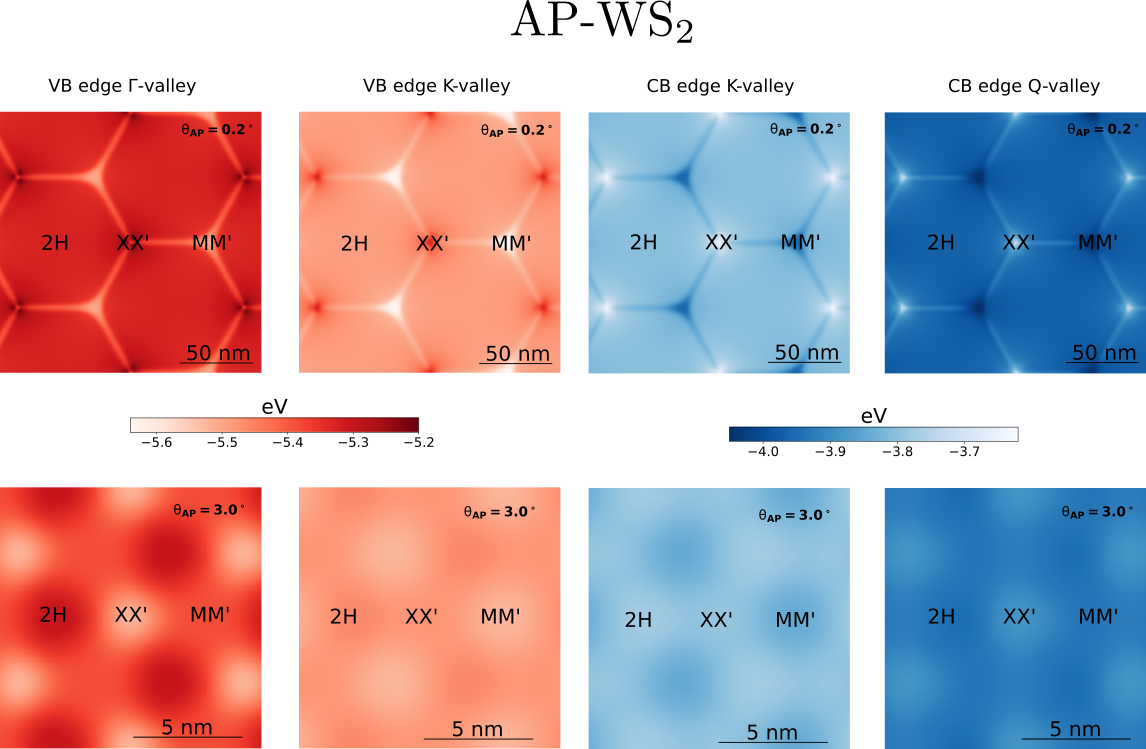}
\caption{Same as in Fig. \ref{fig:APMoS2_band_maps_supl} for AP-WS$_2$}
\label{fig:APWS2_band_maps_supl}
\end{figure}

\begin{figure}[h]
\includegraphics[scale=0.24]{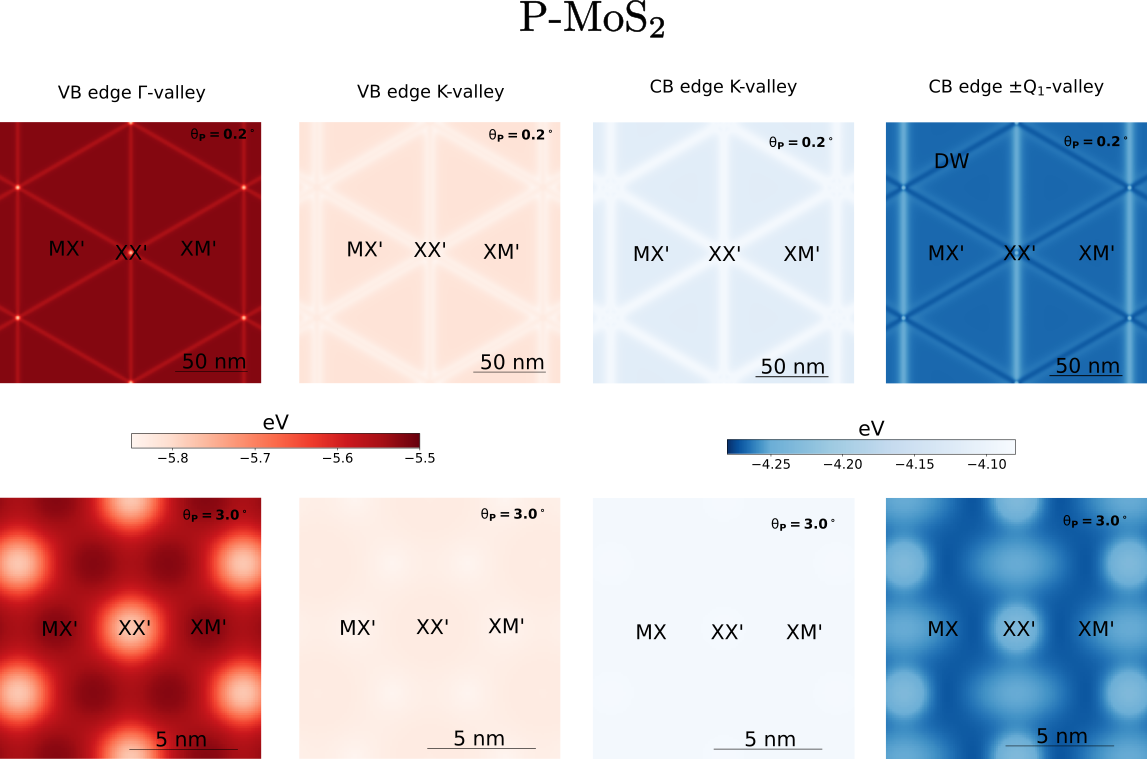}
\caption{Maps of band edge for P-MoS$_2$ homobilayer in the VB at the $\Gamma$- and K-valley and CB at the K and Q$_1$-valley for $\theta_{AP} = 0.2^\circ$ and $\theta_{AP} = 3^\circ$.
See Fig. \ref{fig:stackings_supl} for more details about XX$'$, MX$'$ and XM$'$ stacking configurations.}
\label{fig:PMoS2_band_maps_supl}
\end{figure}

\begin{figure}[h]
\includegraphics[scale=0.24]{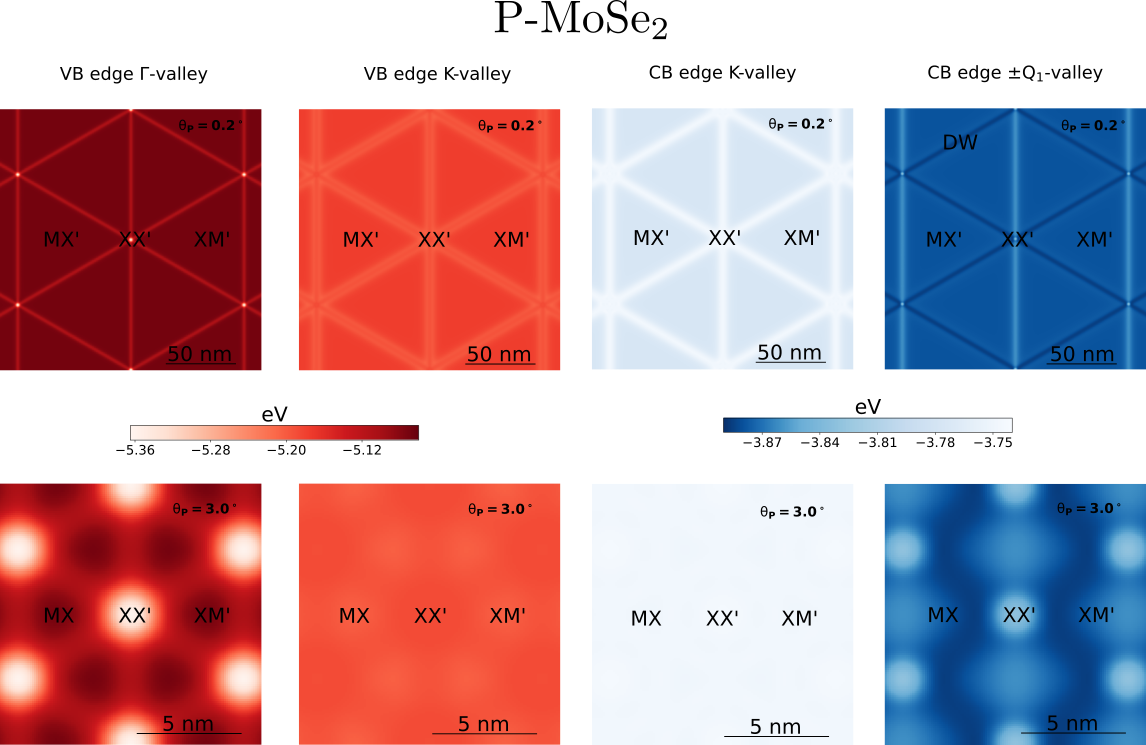}
\caption{Same as in Fig. \ref{fig:PMoS2_band_maps_supl} for P-MoSe$_2$}
\label{fig:PMoSe2_band_maps_supl}
\end{figure}

\begin{figure}[h]
\includegraphics[scale=0.24]{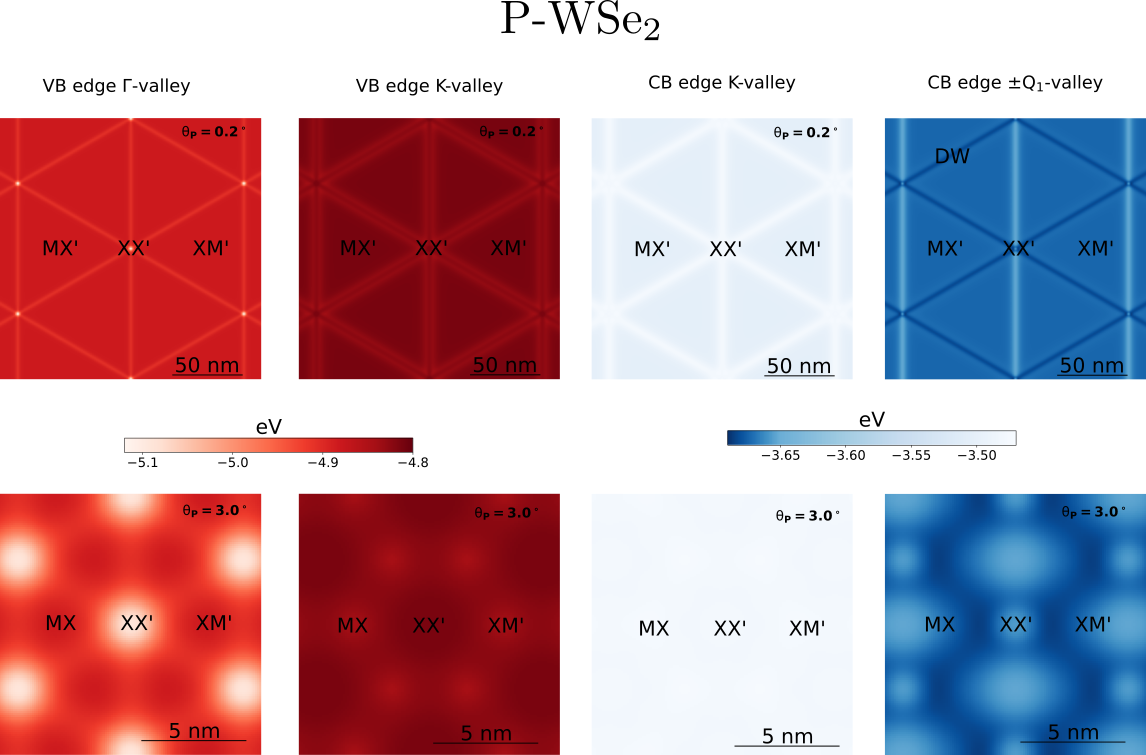}
\caption{Same as in Fig. \ref{fig:PMoS2_band_maps_supl} for P-WSe$_2$}
\label{fig:PWSe2_band_maps_supl}
\end{figure}

\begin{figure}[h]
\includegraphics[scale=0.24]{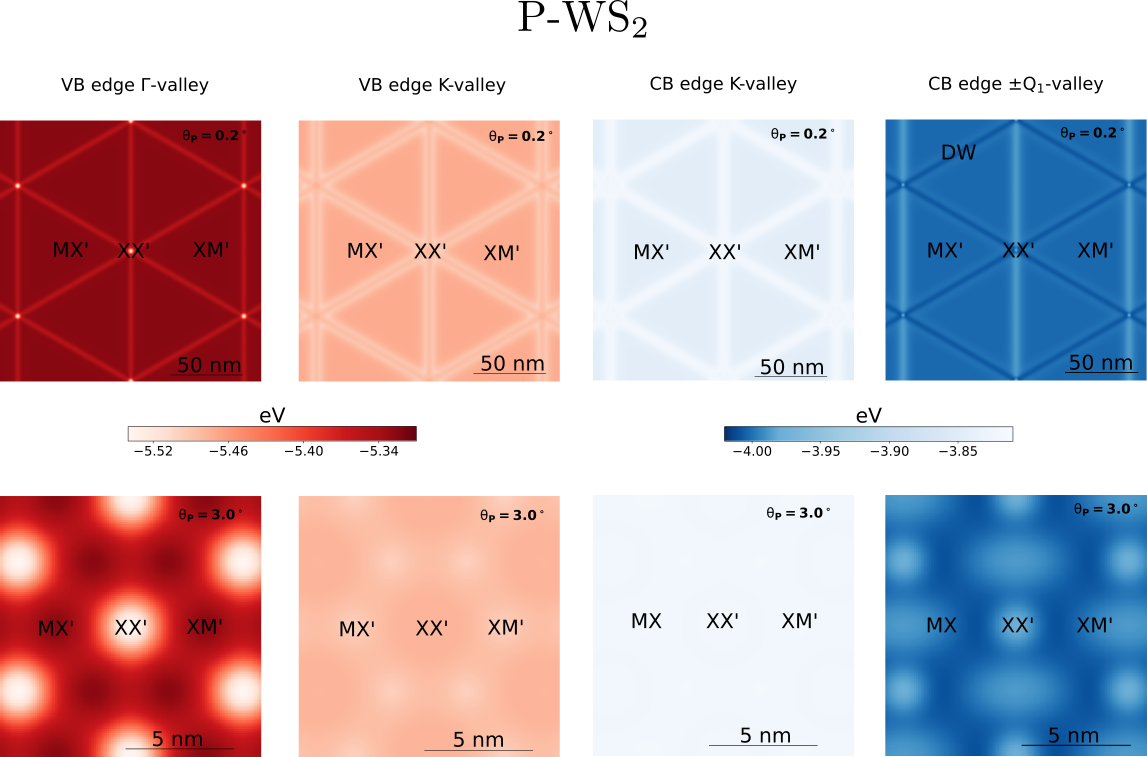}
\caption{Same as in Fig. \ref{fig:PMoS2_band_maps_supl} for P-WS$_2$}
\label{fig:PWS2_band_maps_supl}
\end{figure}

\begin{figure}[h]
\includegraphics[scale=0.24]{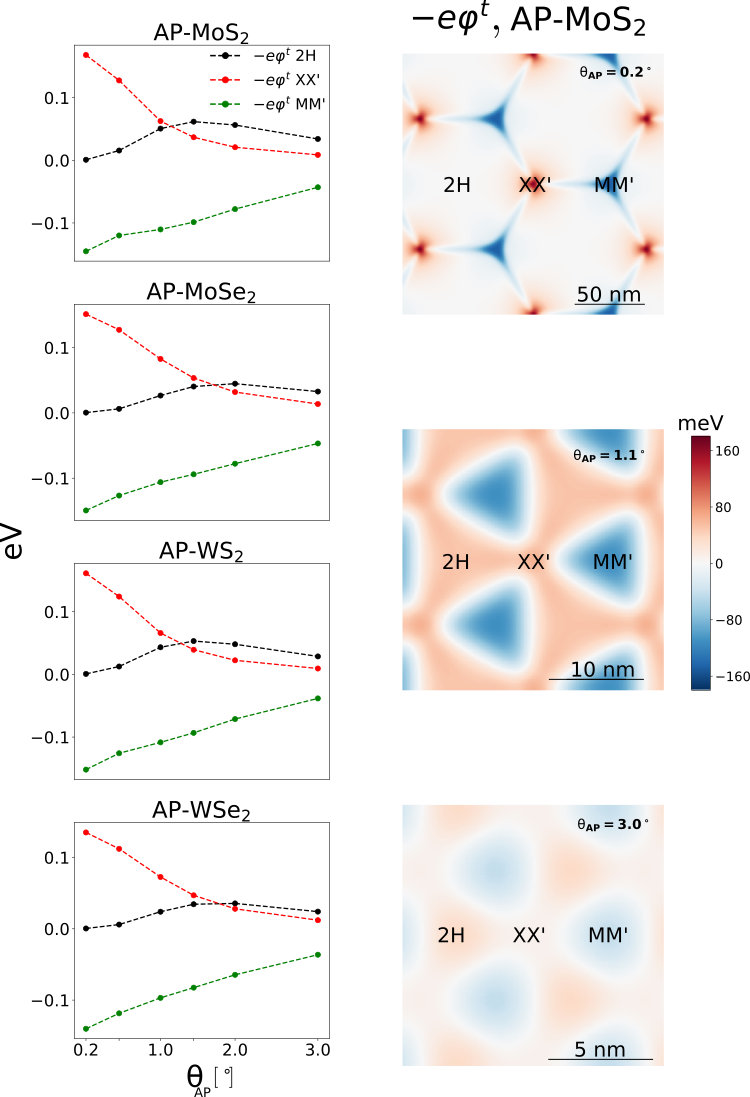}
\caption{Left panels: Variation of piezopotential energy $-e\phi$ for AP-bilayers considered in this study. 
Right panels: Maps of $-e\phi$  for AP-MoS$_2$ with different twist-angles ($\theta_{AP}$).
See Fig. \ref{fig:stackings_supl} for more details about XX, MX$'$ and XM$'$ stacking configurations.}
\label{fig:AP_piezo_maps_supl}
\end{figure}

\begin{figure}[h]
\includegraphics[scale=0.24]{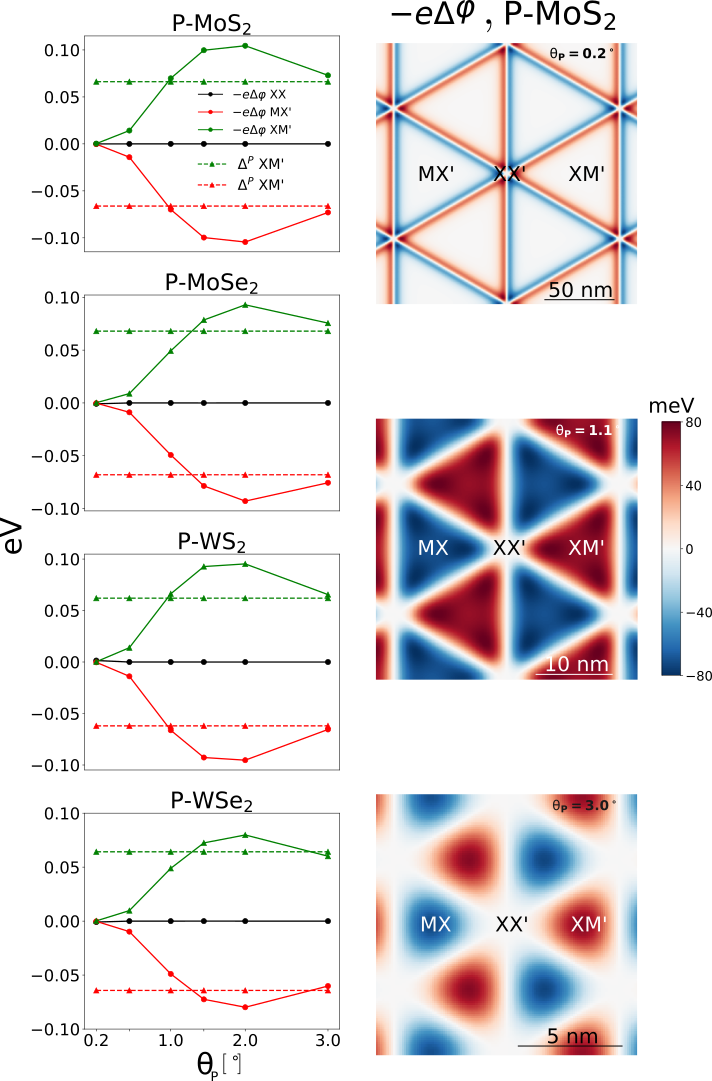}
\caption{Left panels: Map of difference of piezopotentials (in meV) in two layers $-e\Delta \varphi\equiv -2e\varphi$ and ferroelectric potential $\Delta^P$ for P-bilayers considered in this study. 
Right panels: Maps of $-e\Delta \varphi$  for P-MoS$_2$ with different twist-angles ($\theta_P$).
See Fig. \ref{fig:stackings_supl} for more details about XX$'$, MX$'$ and XM$'$ stacking configurations.}
\label{fig:P_piezo_maps_supl}
\end{figure}

To apply the effective Hamiltonians in Eqs. \ref{eq:ef_H_AP_and_P_VB_GAMMA}, \ref{eq:ef_H_P_VB_K}-\ref{eq:ef_H_AP_VB_K}, \ref{eq:ef_H_P_CB_K}-\ref{eq:ef_H_AP_CB_K} and \ref{eq:ef_H_P_CB_Q}, characterising coupling in aligned AP- and P-bilayers, to twisted structures, we substitute $\bm{r}_0(\bm{r})$ in matrix elements by local $\bm{r}_0(\bm{r})$ given by Eq. \ref{eq:local_shiftr0} and local interlayer distance $d$ by $d(\bm{r})=d_0+Z(\bm{r}_0(\bm{r}))$. Finally, we add potential energy of electron in piezopotential (${\rm diag}(-e\varphi,-e\varphi)$ for AP-bilayers and ${\rm diag}(-e\varphi,e\varphi)$ for P-bilayers) to the effective Hamiltonians, where $\varphi\equiv\varphi^t$. Then, we diagonalize resulting Hamiltonians and obtain the following expressions for band edge variation in twisted TMD bilayers:
\begin{equation}\label{eq:BAND_EDGES}
\begin{split}
E_{\rm VB,\Gamma}^{AP}(\bm{r}) = -e\varphi(\bm{r}) + \varepsilon_{\Gamma}(\bm{r}) +  \left|T_{\Gamma}(\bm{r}) + \delta T_{\Gamma}^{AP}(\bm{r})\right|,\\
E^{P}_{\rm VB,\Gamma}(\bm{r})=\varepsilon_{\Gamma}(\bm{r}) + \sqrt{T_{\Gamma}^2(\bm{r}) + \left[\frac{\Delta^P(\bm{r})}{2} +  e\varphi(\bm{r})\right]^2},\\
E_{\rm VB,K}^{AP}(\bm{r})=\varepsilon_{\rm VB,K}^{AP}(\bm{r}) - e\varphi(\bm{r}) + \sqrt{\left|T_{\rm VB, K}^{AP}(\bm{r})\right|^2 +\frac{\left(\Delta^{SO}_{\rm VB, K}\right)^2}{4} },\\
E_{\rm CB,K}^{AP}(\bm{r})=\varepsilon_{\rm CB, K}^{AP}(\bm{r}) - e\varphi(\bm{r}) - \sqrt{\left|T_{\rm CB,K}^{AP}(\bm{r})\right|^2 +\frac{\left(\Delta^{SO}_{CB}\right)^2}{4} },\\
E^{P}_{{\rm VB},{\rm K}}(\bm{r}) = \varepsilon_{\rm VB,K}^P(\bm{r}) +\sqrt{\left|T_{\rm VB, K}^P(\bm{r})\right|^2 + \left[\frac{\Delta^P(\bm{r})}{2} +  e\varphi(\bm{r})\right]^2},\\
E^{P}_{{\rm CB },{\rm K}}(\bm{r}) = \varepsilon_{\rm CB, K}^P(\bm{r}) +\sqrt{\left|T_{\rm CB,K}^P(\bm{r})\right|^2 + \left[\frac{\Delta^P(\bm{r})}{2} +  e\varphi(\bm{r})\right]^2},\\
E^{AP}_{{\rm CB,Q_{1}}}(\bm{r})=\varepsilon_{\rm CB,\mathrm{Q}_1}^{AP}(\bm{r}) - e\varphi(\bm{r}) - \sqrt{\left|T_{\rm Q_{1} }^{AP}(\bm{r})\right|^2 +\frac{\left(\Delta^{\rm Q}_{SO}\right)^2}{4} },\\
E^{P}_{{\rm CB,Q_{1}}}(\bm{r}) = \varepsilon_{\rm CB,\mathrm{Q}_1}^{P}(\bm{r}) +\sqrt{\left|T_{\rm Q_{1} }^P(\bm{r})\right|^2 + \left[\frac{S^P(\bm{r})}{2} +  e\varphi(\bm{r})\right]^2},
\end{split}
\end{equation}
In Figs. \ref{fig:APMoS2_band_maps_supl}-\ref{fig:PWS2_band_maps_supl} are shown band edge maps of homobilayers considered in this work for twist-angles of $0.2^\circ$ and $3^\circ$. 
Maps of piezopotentials across the moir\'e supercell for AP- and P-MoS$_2$ are shown in Figs. \ref{fig:AP_piezo_maps_supl}-\ref{fig:P_piezo_maps_supl}.
In Table \ref{tab:domiant_contributions} are summarized the dominant effects that determine the location of the band edge in the valence and conduction bands for TMD homobilayers under consideration for marginal twist angles.
Interlayer hybridization effect is strong at $\Gamma$-valley for both P- and AP-bilayers.
This is due to the contribution from $d_z^2$ and $p_z$ atomic orbitals of metals and chalcogens respectively, that makes band edge value sensitive to interlayer distance variation. 
Piezoelectric potential plays an important role around corners of 2H in AP-bilayers for marginal angles. 
This promotes the band edge at $\Gamma$-valley to be located in corners of 2H domains. 
On the other hand,  the piezoelectric potential is not as important for P-bilayers, making interlayer hybridization as the only important term to determine the band edge location, in this case at MX$'$ stacking.
In the valence band at the K-valley, the lack of contribution from atomic orbtials of chalcogens makes it not so sensitive to variation of interlayer distance. 
Thus, the interlayer hybridization contribution does not vary significantly across the moir\'e supercell.
For AP-bilayers, we see that the variation of the band edge at K-valley with twist angle for valence and conduction bands depends on the magnitude of the piezopotential. 
Therefore, piezoelectricity is the dominant effect for such a valley promoting the band edge to be located at XX$'$ and MM$'$ for the valence and conduction bands respectively at the K-valley.
Ferroelectricity in P-bilayers is another important effect that influences the location of the band edge.
The only TMD considered in this study where the interlayer hybridization effect is important at K-valley is P-WSe$_2$.
The band edge location at Q in AP-bilayers for marginal angles is mainly affected by the piezopotential.
Just like in the case for the K-valley, piezoelectricity promotes the band edge to be located at MM$'$ for conduction bands. 
For P-bilayers, the band edge location at Q is mostly determined by the interlayer hybridization where ferroelectricity and piezopotential show a negligible contribution to MX$'$/XM$'$ regions. 
The CB minimum is located  along  one-dimensional  channels  zig-zagging across the moir\'e superlattice.
This value does not vary with the twist angle, which shows that interlayer hybridization is the dominant term. 

\begin{figure}[h]
\includegraphics[scale=0.45]{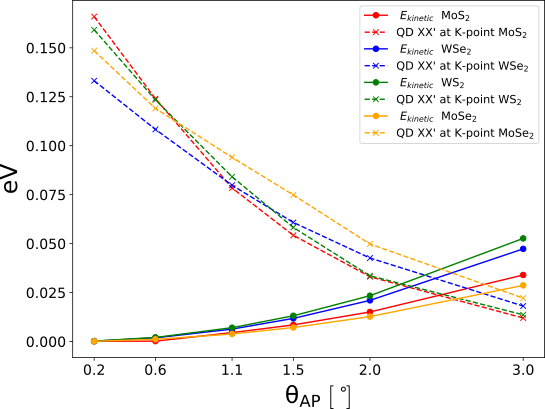}
\caption{Dashed lines represent the depth of quantum dots in XX$'$ regions quantum for  K-valley holes in AP-bilayers. Solid lines represent maximal kinetic energy of the hole given by $E_{kinetic} = (\Delta \rm K)^2 /2m_{\rm K}^*$, where $(\Delta \mathrm{K})^2 = 4 \pi \theta_{AP}^2/3a$ and $m_{\rm K}^*$ is the effective mass of the hole at K taken from Ref. \cite{Korm_nyos_2015}. 
Narrow bands are formed in the regime where quantum dot potential depth is much larger than the kinetic energy. This condition is satisfied in the interval  $\theta_{AP}<1^\circ$.   }
\label{fig:QD_picture}
\end{figure}

Lastly, in Fig. \ref{fig:QD_picture} is shown a comparison of quantum dot potential depth in XX' region for K-valley holes\cite{Korm_nyos_2015} with their maximal kinetic energy in mini Brillouin zone for given twist angle.
Regime of weakly coupled quantum dots is realized when the depth is much larger than the kinetic energy, which leads to the following estimate for twist angles: $\theta_{AP} < 1^\circ$.

\begin{table}[h]
\caption{Dominant effects that determine  the location of the VB and CB edge at marginal twist angles for the TMDs bilayers.
PE, FE and iH  stand for piezoelectricity, ferroelectricity and interlayer hybridization.}
    \begin{tabular}{|c|c|c|c|c|c|c|c|c|}
    \hline
    \multirow{2}{*}{} & \multicolumn{4}{c|}{Parallel} & \multicolumn{4}{c|}{Anti-Parallel} \\ \cline{2-9} 
        &  VB K-valley  & CB K-valley  &CB Q-valley  &VB $\Gamma$-valley & VB K-valley & CB K-valley &CB Q-valley &    VB $\Gamma$-valley          \\ \hline
    MoS$_2$             &  PE and FE   & PE and FE  & iH & iH&  PE & PE  & PE  &iH and PE       \\
    MoSe$_2$            &  PE and FE   & PE and FE  & iH & iH&  PE & PE  & PE  &iH and PE      \\ 
    WS$_2$              &  PE and FE   & PE and FE  & iH & iH&  PE & PE  & PE &iH and PE       \\ 
    WSe$_2$         &  iH, PE and FE   & PE and FE  & iH & iH&  PE & PE  & PE &iH and PE       \\ \hline
    \end{tabular}
\label{tab:domiant_contributions} 
\end{table}

\twocolumngrid

\end{document}